\documentclass[preprint2]{aastex}
\usepackage{epsfig}
\usepackage{natbib}

\newcounter{subfigure}

\begin{document}
\newcommand{\kms}{km~s$^{-1}$}
\newcommand{\Msun}{M_{\odot}}
\newcommand{\Lsun}{L_{\odot}}
\newcommand{\ML}{M_{\odot}/L_{\odot}}
\newcommand{\etal}{{et al.}\ }
\newcommand{\hhh}{h_{100}}
\newcommand{\hsq}{h_{100}^{-2}}
\newcommand{\tn}{\tablenotemark}
\newcommand{\mdot}{\dot{M}}
\newcommand{\p}{^\prime}
\newcommand{\kmsMpc}{km~s$^{-1}$~Mpc$^{-1}$}

\title{Cosmicflows-2: I-band Luminosity - HI Linewidth Calibration}

\author{R. Brent Tully,}
\affil{Institute for Astronomy, University of Hawaii, 2680 Woodlawn Drive, Honolulu, HI 96822, USA}

\and
\author{H\'el\`ene M. Courtois}
\affil{Institute for Astronomy, University of Hawaii, 2680 Woodlawn Drive, Honolulu, HI 96822, USA\
University of Lyon; UCB Lyon 1/CNRS/IN2P3/INSU; Lyon, France}

\begin{abstract}
In order to measure distances with minimal systematics using the correlation between galaxy luminosities and rotation rates it is necessary to adhere to a strict and tested recipe.  We now derive a measure of rotation from a new characterization of the width of a neutral Hydrogen line profile.  Additionally, new photometry and zero point calibration data are available.  Particularly the introduction of a new linewidth parameter necessitates the reconstruction and absolute calibration of the luminosity-linewidth template.  The slope of the new template is set by 267 galaxies in 13 clusters.  The zero point is set by 36 galaxies with Cepheid or Tip of the Red Giant Branch distances.  Tentatively, we determine H$_0 \sim 75$~\kmsMpc.  Distances determined using the luminosity-linewidth calibration will contribute to the distance compendium Cosmicflows-2.
\end{abstract}

\
\section{Introduction}

Cosmicflows-1 is the re-christained name of the catalog of galaxy distances published by \citet{2008ApJ...676..184T} and available at the Extragalactic Distance Database (EDD)\footnote{http://edd.ifa.hawaii.edu}.  That catalog contains distance measures for 1797 galaxies, in the majority from the correlation between galaxy luminosity and rotation rate \citep{1977A&A....54..661T} but also including Cepheid Period-Luminosity Relation \citep{2001ApJ...553...47F}, Tip of the Red Giant Branch \citep{2007ApJ...661..815R, 2009AJ....138..332J}, and Surface Brightness Fluctuation \citep{2001ApJ...546..681T} distance estimates.  Cosmicflows-1 is quite limited in depth, extending to only 3,000~\kms.  However it provides the densest coverage of the historical Local Supercluster of any available catalog.

Cosmicflows-2 is a catalog in construction that will build on the high density of local coverage but will extend to considerably greater distances.  In addition to an update of the Cepheid, Red Giant Branch, and Surface Brightness Fluctuation information the new catalog will include Type Ia Supernova \citep{2007ApJ...659..122J} and Fundamental Plane \citep{2001MNRAS.321..277C, 2001MNRAS.327..265H, 2002AJ....123.2990B} material.  However, once again, the numerically dominant source of distances will come from the luminosity - HI linewidth correlation.  \citet{2011MNRAS.414.2005C} have described several luminosity-linewidth samples that will be major components of Cosmicflows-2.  The purpose of the present paper is to provide a new calibration of the luminosity-linewidth correlation.

The calibration of this correlation used with Cosmicflows-1 was described in
detail by \citet{2000ApJ...533..744T}, with minor updates in \citet{2008ApJ...676..184T}.  The linewidth measure used with those earlier discussions is $W_{20}$, the width of an HI profile at 20\% of peak intensity.
This is an analog measure with origins that date back to the first paper by
Tully \& Fisher.  As we move forward with digital data and incorporate larger
samples it is an appropriate moment to reconsider the details of the
measurement of a profile width.  The new profile measure that we choose is different enough that it imposes the need for a new end-to-end re-calibration of the methodology.

While the recent influx of high quality digital HI profile information is the largest driver for this re-calibration, there is considerably more photometric material available today and more galaxies are available with accurate Cepheid or Tip of the Red Giant Branch distance determinations that can establish the scale zero point.  This combination of advances motivates the re-calibration to be discussed as a prelude to the publication of Cosmicflows-2.

\section{Calibrator Sample}

The calibration of the luminosity-linewidth correlation involves two important
steps; the first with the establishment of the slope of the correlation that
is critical for the minimization of a form of Malmquist bias and the second
with the establishment of the scale zero point.    The slope of the
correlation is given through an analysis of samples drawn from clusters of
galaxies.  The details of the choice of clusters are discussed in Section~\ref{sec:template} and given elaboration in the Appendix.  In the present study the correlation slope is given by data from 267 galaxies drawn from 13 clusters, compared with 241 galaxies drawn from 12
clusters used earlier \citep{2000ApJ...533..744T}.   With the slope fixed, the scale zero point is established from the luminosity-linewidth properties of relatively nearby galaxies with independently determined distances.  This problem is discussed in Section~\ref{sec:zp}.
Presently the scale zero point is set by 36 galaxies with well determined Cepheid Period-Luminosity \citep{2001ApJ...553...47F} or Tip of the Red Giant Branch \citep{2007ApJ...661..815R} distance measures, compared with 24 available earlier.

Implementation of the luminosity-linewidth methodology requires the observation of three parameters: (1) a measure of the rotation rate obtained by observing the width of the 21cm HI spectral line, (2) the apparent luminosity of the galaxy, in this case represented by the flux in the near-infrared $I$ band, and (3) the inclination of the galaxy, inferred from the ellipticity of the photometric image, and used to de-project the measure of rotation to what would be seen in an edge-on view and to de-project reddening to what would be seen in a face-on view.   The relevant data are being accumulated for the large Cosmicflows-2 sample.  The following two subsections summarize the transformation steps from raw observables to parameters used in fitting for distances. 

\subsection{HI Linewidths}

Major radio observatories have preserved digital HI spectral flux data from observations extending back more than two decades.    We have gathered over 12,000 profiles from these archival sources and complemented them with some 2000 profiles from our own recent observations.  The task we then undertook was to analyze all this material in a consistent way.

An HI profile linewidth that we might measure is a utilitarian parameter only indirectly connected with the physics of galactic rotation.  Several alternative linewidth definitions were explored by \citet{2005ApJS..160..149S}.  We have settled on a variant of one of these.  The justifications and details of our preferred linewidth measure are given by \citet{2009AJ....138.1938C, 2011MNRAS.414.2005C}.  
In brief, we want a measure with the following properties: (1) at a level low enough to capture the full range of rotation motions while high enough to usually be above noise and wings in profiles that are adequately observed and (2) not sensitive to details of the profile shape; ie, whether the profile is single or double peaked or whether the flux in peaks are asymmetric.   
The measured parameter is $W_{m50}$, the linewidth at 50\% of the mean flux per channel contained within spectral channels embracing 90\% of the total flux.  Width error estimates are based on the level of the signal, $S$, at 50\% of mean flux over the noise, $N$, measured beyond the extremities of the signal \citep{2009AJ....138.1938C}.  Useful profiles have error estimates no greater than 20~\kms\ corresponding to a flux per channel $S/N \ge 2$.  This minimum $S/N$ level corresponds approximately to a {\it peak} signal to noise ratio of seven.

It was described by Courtois et al. that the linewidth parameter must be corrected for redshift and instrumental broadening
\begin{equation}
W_{m50}^c = {{W_{m50}} \over {1+z}} - 2 \Delta v \lambda
\label{Wc}
\end{equation}
where $cz$ is the heliocentric velocity of the galaxy, $\Delta v$ is the spectral resolution after smoothing, and $\lambda=0.25$ is determined empirically.   There was further discussion that this utilitarian parameter could be transformed into an approximation to a physically meaningful parameter: the velocity difference between the positive and negative extremes of rotation in the galaxy.  The physically motivated parameter is $W^i_{mx}$:
\begin{eqnarray}
\nonumber
W_{mx}^2  = W_{m50}^2 + W_{t,m50}^2 [1 - 2  e^{-(W_{m50}/W_{c,m50})^2}] \\
 -  2 W_{m50} W_{t,m50} [ 1 - e^{-(W_{m50}/W_{c,m50})^2}]  
\label{WR}
\end{eqnarray}
with $W_{c,m50} = 100$~\kms\ descriptive of the transition from box-car to gaussian profile shapes and $W_{t,m50} = 9$~\kms\ a parameter determined by random internal motions.  Finally, a projection correction is made with $W^i_{mx} = W_{mx}/{\rm sin}(i)$ where $i$ is the galaxy inclination from face on.  This parameter $W^i_{mx}$ that approximates twice the maximum rotation velocity of a galaxy is the new linewidth parameter to be used in the correlation with luminosities, replacing the parameter $W^i_R$ introduced by 
\citet{1985ApJS...58...67T}.

All the HI profiles used in the current analysis can be seen at the web site http://edd.ifa.hawaii.edu by selecting the catalog `All Digital HI' then searching for the desired galaxy in that catalog and clicking on the name in the second column.

\subsection{I-band Photometry}

Photometry continues to accumulate, with particular note to the products by the Cornell group \citep{2006ApJ...653..861M}.  Our own recent contribution was reported by 
\citet{2011MNRAS.tmp..796C} where attention was given to assure that the photometry in an approximation to the Cousins $I$-band is consistent between authors.

Luminosities must be corrected for extinction internal to the target galaxies and our own Galaxy and for the displacement of flux due to Doppler shift.  Extinction internal to the targets is usually the greatest concern.    We follow the prescription given by \citet{1998AJ....115.2264T} who found that internal extinction has a pronounced luminosity dependence that can be described in terms of the rotation parameter.  The inclination dependent internal extinction at $I$-band is described by the expression $A^I_i = \gamma_I {\rm log} (a/b)$ where $a/b$ is the major to minor axis ratio, $i$ is the inclination defined by the expression ${\rm cos}~i = [((b/a)^2 -q_0^2)/(1-q_0^2)]^{1/2}$ with $q_0 = 0.20$ the statistical axial ratio of a galaxy viewed edge-on, and $\gamma_I$ has the specification
\begin{equation}
\gamma_I = 0.92 +1.63 ({\rm log} W^i_{mx} - 2.5).
\label{gamma}
\end{equation}
Arguments can be made for a more complex dependence of $q_0$ but we justify in \citet{2000ApJ...533..744T} our simplified choice.

Galactic extinction is important at low latitudes.  We follow \citet{1998ApJ...500..525S}, using the correction term $A^I_b = R_I E(B-V)$ with differential reddening $E(B-V)$ determined from $100~\mu$m cirrus maps and $R_I = 1.77$.  The third adjustment, the $k$-correction, is very minor for the small redshifts we encounter.  \citet{2010MNRAS.405.1409C} recommend a correction which is more than adequate, $A^I_k = 0.302 z + 8.768 z^2 -68.680 z^3 +181.904 z^4$.  A fully corrected magnitude is
\begin{equation}
 I_T^{b,i,k} = I_T -A_b^I -A_i^I-A_k^I.
 \label{Icor}
 \end{equation}

\section{Nulling Bias}

Our concern in this discussion is with one of two distinct `Malmquist biases'  (Malmquist 1920, 1922, 1924).  Confusion has arisen with Malmquist bias terminology so, to be clear, we are not at this point considering the bias that results from the volume sampling effect: either the first order effect with a `homogeneous' distribution where a galaxy with a given measured distance is more likely to be more distant and scattered inward by error than nearer and scattered outward \citep{1988ApJ...326...19L}, or the second order effect with an `inhomogeneous' distribution where errors scatter measures out of higher into lower density regions \citep{1995PhR...261..271S}.  With the volume sampling Malmquist biases, individual distance measures may statistically scatter symmetrically about correct distances yet a map of peculiar velocities might be biased, for example in the case of the homogeneous distribution case,  due to the preponderance of inward over outward scattered distance measures since a more distant shell statistically contains more galaxies than a nearer shell of the same thickness.  This potential bias can be minimized but the matter is best considered in the context of a velocity field analysis \citep{1995ApJ...454...15S}.

The Malmquist effect that concerns us here is the one induced by a magnitude
cutoff that can bias individual distance measures.\footnote{\citet{1995PhR...261..271S} call this a selection effect rather than a Malmquist effect.}  Considering two galaxies
with the same linewidth and at the same distance, the fainter one may be lost
from the sample while the brighter one is retained \citep{1984A&A...141..407T,
  1994ApJ...430...13S, 2007A&A...465...71T}.  Intuitively it will be expected that if galaxies
intrinsically brighter than the mean tend to be selected, but are attributed
the mean intrinsic luminosity, then galaxies will tend to be placed too close.
This is the Malmquist bias.  The amplitude of the bias depends on the slope
attributed to the correlation.  Intuition can be compromised; the bias can
actually be reversed in sign by selecting a very steep slope because galaxies
will tend to lie below the posited relation at the bright, high linewidth end.  It follows that there is a slope that neutralizes the bias.  This slope is given by a fit to the luminosity-linewidth correlation that assumes errors are entirely in linewidths, the `inverse TF relation' \citep{1980AJ.....85..801S, 1988Natur.334..209T, 2000ApJ...533..744T}.

A critical step in what follows is the construction of a template that gives the proper slope for bias-free distance measures.  The template might be constructed from a cluster sample where all the galaxies are at the same distance and all appropriate galaxies are included down to a defined faint level.   With a least squares fit assuming errors are in linewidth, galaxies in each magnitude interval will scatter symmetrically about the fit.  If this template relation is used to determine the distance to a galaxy in the field, the linewidth of that target galaxy will be drawn from the same symmetric distribution about the mean, as likely to be on one side of the relation in linewidth at a given magnitude as on the other.  The distance measure will be unbiased.

This procedure involves several assumptions.  It is assumed that the luminosity-linewidth correlation obeys a power law and that it is universal at the current epoch. It assumes that the scatter, which is both intrinsic and observational, is similar with both the template and the field targets.  The procedure requires that the sample not be restricted in the linewidth measurements because any restrictions would carry the bias over to the orthogonal axis.   The potential for bias depends on the luminosity function of normal gas-rich galaxies since if there are increasing numbers at fainter intervals then errors scatter more galaxies upward in luminosity than down.  Moreover, sample selections have not been made with strict $I$ band limits, rather, historically selections have been made from catalogs such as the Catalogue of Galaxies and Clusters of Galaxies \citep{1968cgcg.bookR....Z} or Uppsala General Catalogue \citep{1973ugcg.book.....N} of the northern sky or the ESO/Uppsala Atlas \citep{1982euse.book.....L} of the southern sky.  In these cases, selections are by photographic magnitude or diameter and transformations to $I$ band involve color or surface brightness dependencies.  \citet{1994ApJS...92....1W} conducted a particularly thorough analysis of biases.  He is a bit misleading in the abstract of that paper with the following statement.

\noindent{\it 
"The use of the inverse method has been seen by some workers as a panacea for bias effects.  A goal of this series is to show that in general the inverse method exhibits biases analogous to, though different in detail from those exhibited by the forward relation."}

Willick goes on to show that while there can be bias with the inverse relation it is much reduced over the forward relation.  In a concrete example \citep{1995ApJ...446...12W} the bias correction is reduced by a factor 6 from the forward relation, reducing the bias from a significant concern to a marginal effect.  \citet{2006ApJ...653..861M}, and before that \citet{1997AJ....113...53G}, have discussed the bias in the case of a bivariate formulation of the luminosity-linewidth relation where the bias is intermediate in amplitude.  We will describe tests with simulations to estimate the level of bias pertinent to our procedures.  

\begin{figure}[t]
\centering
\includegraphics[scale=0.39]{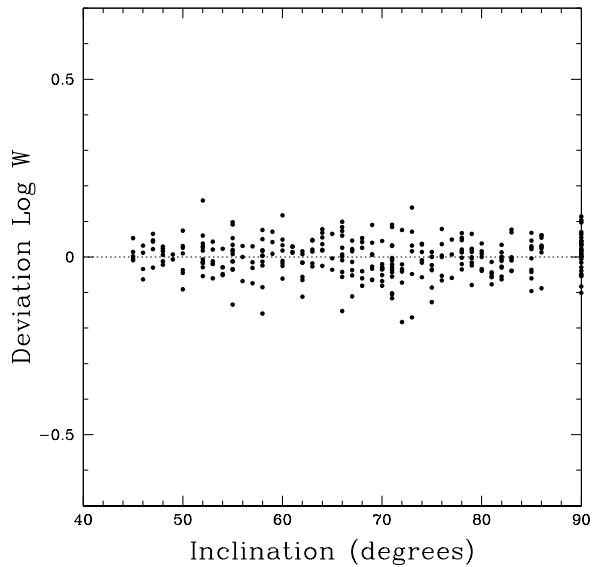}
\caption{Deviations in Log $W_{mx}^i$ from the mean luminosity-linewidth relation as a function of inclination.}
\label{tfi0}
\end{figure}

\begin{figure}[!]
\centering
\includegraphics[scale=0.39]{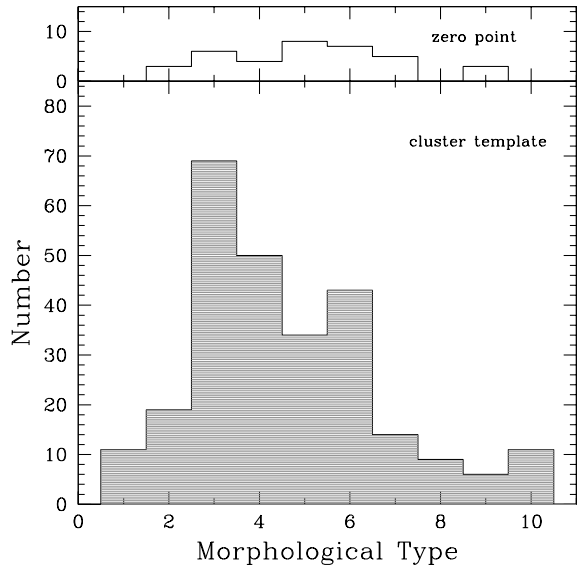}
\caption{Distribution of morphological T types for the galaxies in the 13 cluster template sample (filled lower histogram) and for the zero point calibrators (open upper histogram).} 
\label{types}
\end{figure}

Preliminary to that, we review properties of our sample.
(1) Galaxies judged to be more face-on than $45^{\circ}$ are excluded.   Tests have not shown there to be any inclination dependencies in distance measures for samples restricted to this limit.  Figure~\ref{tfi0} shows that with our cumulative sample there are no systematics as a function of inclination in deviations from the mean relationship that is ultimately derived. There is not even any increase in scatter either toward $45^{\circ}$ where linewidth projection uncertainties are greatest or toward $90^{\circ}$ where absorption corrections are greatest.  (2) Galaxies earlier in type than Sa are excluded.  If S0 galaxies are accepted there is a substantial increase in scatter.  Sa are on the cusp; sometimes they scatter badly.  Exclusion on the basis of morphology introduces an unfortunate qualitative aspect. The distribution of types used in our calibration is illustrated in Figure~\ref{types}.  Only 11 cases (4\%) of the template calibrators are typed T=1 (Sa).  It is to be noted that the sample is magnitude but not type limited on the faint end (irregulars are included if they are bright enough).  The template sample is weighted toward earlier spiral types because many of the template clusters are sufficiently distant that late types are excluded by the apparent magnitude limit. (3) Galaxies are excluded if they are pathological, show pronounced evidence of tidal interactions, or have confused HI profiles.  We tend to be inclusive if distortions are minor but contributions to the HI flux from multiple sources is a clear basis for exclusion. (4) Galaxies cannot be included if their HI profiles have inadequate signal to noise. The threshold requirement was discussed in Section 2.1.  This restriction can be serious if it affects preferential parts of the correlation, for example, large linewidth systems where the flux per channel is reduced.  We have addressed this issue by using long integrations to try to assure adequate coverage of calibrators.  Galaxies too deficient to be reasonably detected are usually typed S0-Sa and are subject to elimination on the basis of morphology. It is to be noted that we do not impose a lower limit on linewidths nor do we reject based on line shape (eg, single vs. double peaked).

\subsection{Bias in Simulations}

\begin{figure}[t]
\centering
\includegraphics[scale=0.39]{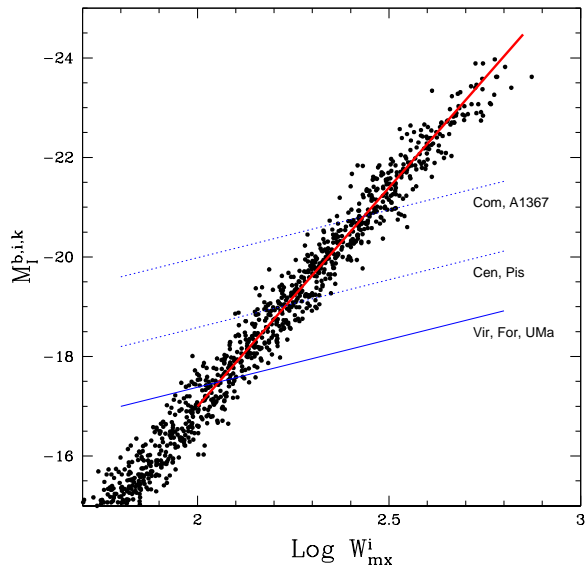}
\caption{Simulated luminosity-linewidth correlation consistent with a luminosity function with $M_I^{\star}=-23$, slope $\alpha = -1.0$, scatter $\sigma_m = 0.4$, and a correlation inverse slope and zero point of $-8.81$ and $-21.39$, the same as the observed relation shown by the heavy red line.  The solid slanting line illustrates the $I$ band magnitude cutoff for the nearest 3 clusters.  Dotted slanting lines show representative cutoffs for more distant clusters.} 
\label{sim}
\end{figure}

While Malmquist selection biases are expected to be small using the inverse correlation  \citet{1994ApJS...92....1W} warns that they are not zero and the effect is systematic.  We have run simulations to evaluate the amplitude of the problem.  It was first established that the combined Virgo$-$Fornax$-$Ursa Major sample entailing 75 galaxies above an absolute magnitude limit $M_I = -17.5$ are described by a luminosity function \citep{1976ApJ...203..297S} with bright end exponential cutoff parameter $M_I^{\star} = -23.0$ and faint end power law slope parameter $\alpha = -1.0$.  A simulated luminosity-linewidth correlation was created by randomly drawing luminosities in accordance with the observed Virgo$-$Fornax$-$Ursa Major luminosity function and associating linewidths with a prescription that gave the desired correlation.  Scatter to the relation was introduced by randomly drawing from a normal distribution to generate a dispersion in luminosity of 0.4 mag.  A second simulated sample was created with scatter generated in linewidth with amplitude that translated on the luminosity axis to 0.4 mag.  In each case, 1000 points were simulated.  The sample with scatter in linewidth is shown in Figure~\ref{sim}.  The simulation was extended down to $M_I=-15$, much lower than the useful observational limit, to ensure proper sampling at the faint limit.  Iterations of the luminosity-linewidth slope were required to produce a correlation with the proper inverse slope of $-8.81$ and scatter of 0.4 mag.  Convergence was achieved with an input bivariate slope of $-8.56$, consistent with a relationship between luminosity and rotation of $L_I \propto W^{3.42}$.

\begin{figure}[t]
\centering
\includegraphics[scale=0.39]{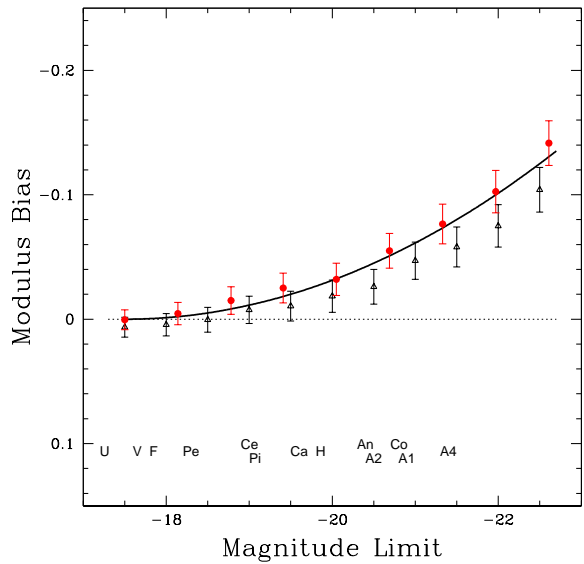}
\caption{Bias as a function of cutoff magnitude.  Open triangles: flat magnitude cutoff.  Filled circles and solid curve: slanted magnitude cutoff at $I$ band.  Error bars are determined by repetitions varying the random number generator seed. Cutoffs for the 13 clusters are indicated by the abbreviated names above the bottom axis.}
\label{bias}
\end{figure}

Differentials in magnitude with respect to the thick solid line in Figure~\ref{sim} (the inverse luminosity-linewidth relation) correspond to differentials in measured `distance moduli', with points above the line given distances closer than the mean and points below the line given distance farther than the mean.  Bias arises from two effects.  The more important one is generated by the form of the luminosity function, particularly the bright end attenuation.  If the luminosity function invokes equal numbers per magnitude interval (our case at the faint end) then scatter upward and downward would be statistically equal and there would be no bias with a magnitude constant cut.  However due to the bright end attenuation fewer points are available to scatter down than up.  Averaged over the ensemble to a specified magnitude limit, the result is an excess of points above the fiducial relation that is very small if the limit is faint but becomes increasingly important as the limit is moved brightward.  The situation is aggravated if the faint limit cut is not at a constant magnitude but, rather, slopes upward with increasing linewidth as anticipated if selection is made in a blue band or diameter.  Such selection favors inclusion of points above the fiducial relation over points below the fiducial relation.  The slanting lines in Figure~\ref{sim} illustrate limits at fixed $B$ magnitudes with typical $B-I$ colors for spiral galaxies \citep{1998AJ....115.2264T}.  The amplitude of the bias in the simulated sample is shown in Figure~\ref{bias} normalized to zero for the nearest 3 clusters (Virgo, Fornax, and Ursa Major) which are anchored by the zero point calibration.  The bias with flat magnitude limits is shown by the black open triangles and the bias with slanting color terms is shown by the red filled circles.  It is seen that in the latter case the bias is only mildly increased.  The solid curve approximates the bias in this latter case.  It obeys the equation $b = 0.005 (\mu -31)^2$ where $b$ is the bias in distance modulus and $\mu -31$ is the distance modulus normalized so $b=0$ at moduli less than 31 (where sample magnitude limits are fainter than the useful range of the luminosity-linewidth correlation).   The bias looks impressive in Figure~\ref{bias} but it has a small effect on distance measures, reaching 3.5\% in the most extreme case (Abell 400).  Cumulatively, the effect on the distances to the 13 calibrator clusters is 2\%.

\section{Building a Template}
\label{sec:template}

Our creation of a template from cluster samples is similar to the "basket of
clusters" approach taken by the Cornell group \citep{1997AJ....113...22G, 2006ApJ...653..861M,
  2007ApJS..172..599S}.  However there are differences.  The Cornell group
derive a slope from a bivariate fit to their template sample rather than an
`inverse' fit with errors solely in linewidths.  Their procedure requires a
specification of the nature of scatter occurring along two axes, which is
beyond our more relaxed assumption that scatter, whatever the source and the error partition between the axes, is
similar for template and targets.   We use fewer clusters but have more candidates per
cluster.\footnote{Overall the Cornell group incorporates many more galaxies because they use optical rotation rate information \citep{2005AJ....130.1037C, 2007AJ....134..334C} in addition to HI information.} We need a significant number of galaxies per cluster to meaningfully
test the universality of the luminosity-linewidth correlation and then to fit
each contributing cluster to the template.  With our 13 clusters, the median
number of galaxies in our sample per cluster is 17, the mean is 20, and the
maximum is 58.  Finally, we do not provide a description of the correlation
with variations that depend on type   \citep{2006ApJ...653..861M,
  2007ApJS..172..599S} or surface brightness \citep{1998A&A...331....1T}.  We view these
variations as only marginally significant in the inverse correlations and they introduce complexity and subjectivity that may vary with distance.

\begin{figure}[t]
\includegraphics[scale=0.39]{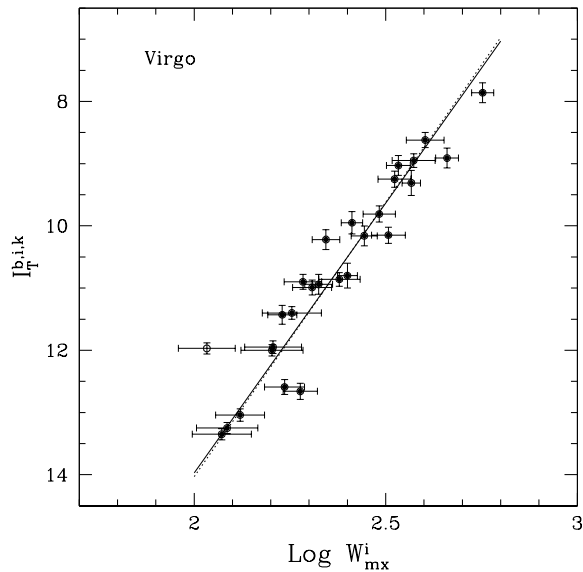}
\caption{$I$ band luminosity $-$ linewidth correlation for 26
  galaxies drawn from the Virgo Cluster. The $I$ band apparent magnitudes are
  corrected for obscuration in our Galaxy ($b$), obscuration in the target
  galaxy ($i$), and the redshift displacement of the spectrum ($k$).  The
  linewidth measure $W_{mx}^i$ is statistically representative of twice the
  maximum rotation velocity including the projection adjustment for
  inclination ($i$).  Dotted line: least squares fit with errors in linewidths
  to the cluster data.  Solid line: universal fit derived from the 13 clusters
  template. The galaxy represented by the open circle was rejected from the fit.  See discussion of Virgo Cluster in Appendix.}
\label{tfivir}
\end{figure}

The first step in the construction of the template involves developing inverse fits (errors in linewidths) to each of the 13 clusters separately.  Results are displayed in Figures~\ref{tfivir}, \ref{tf6a}, and \ref{tf6b}.    

\renewcommand{\thefigure}{\arabic{figure}\alph{subfigure}}
\setcounter{subfigure}{1} 
\onecolumn
\begin{figure}[!]
\includegraphics[scale=0.85]{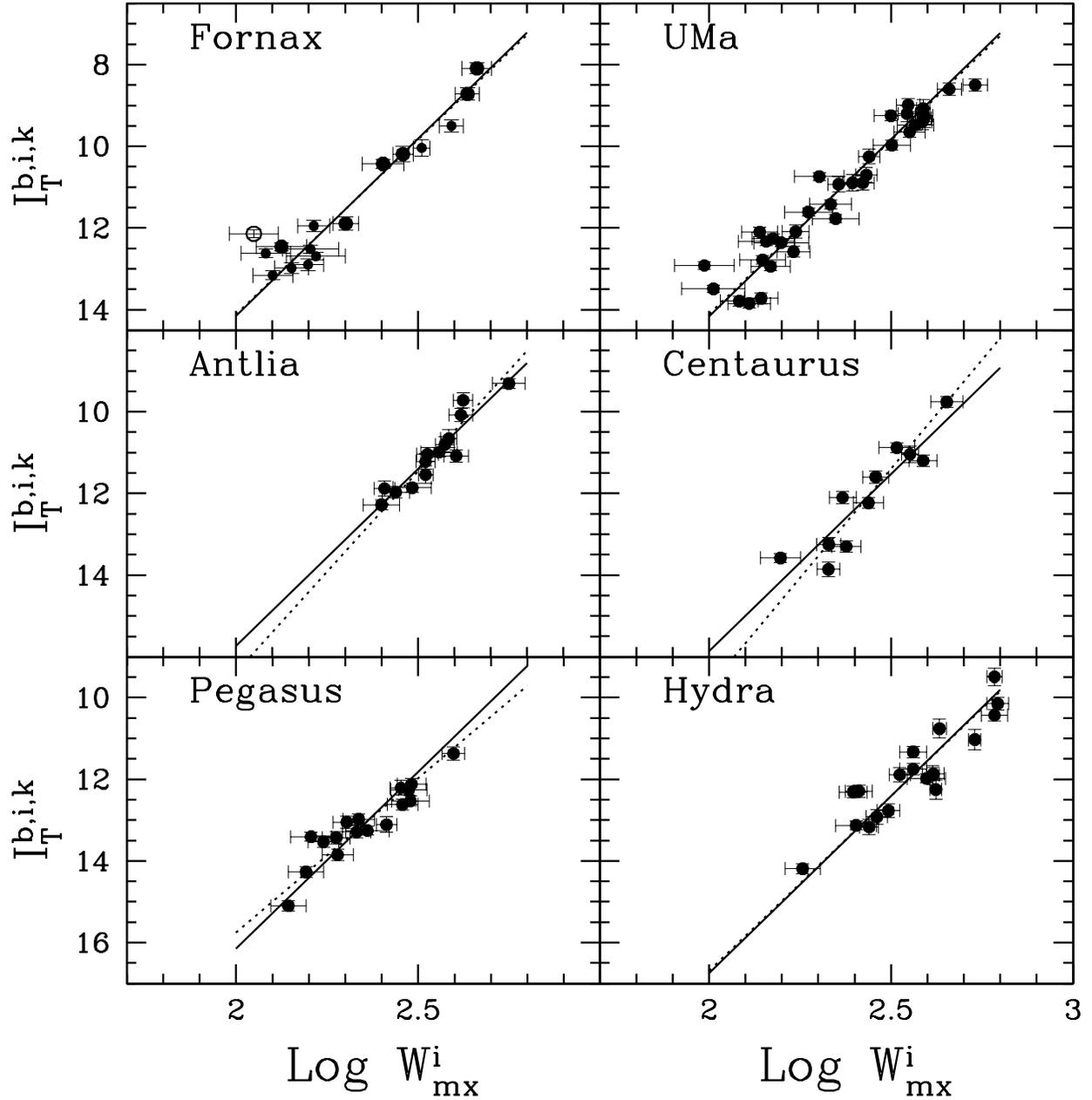}
\caption{$I$ band luminosity$-$linewidth correlations for 15 galaxies in Fornax, 34 galaxies in Ursa Major, 14 galaxies in Antlia, 11 galaxies in Centaurus, 17 galaxies in Pegasus, and 19 galaxies in Hydra.  Solid lines are the universal template fits and dotted lines are fits to the individual clusters.  In the case of the Fornax Cluster, galaxies superimposed on the central core are identified with larger symbols, galaxies in the periphery are identified by smaller symbols, and one rejected galaxy is identified by an open symbol.  See Appendix for discussions of individual clusters.}
\label{tf6a}
\end{figure}

\addtocounter{figure}{-1}
\addtocounter{subfigure}{1}
\begin{figure}
\includegraphics[scale=0.85]{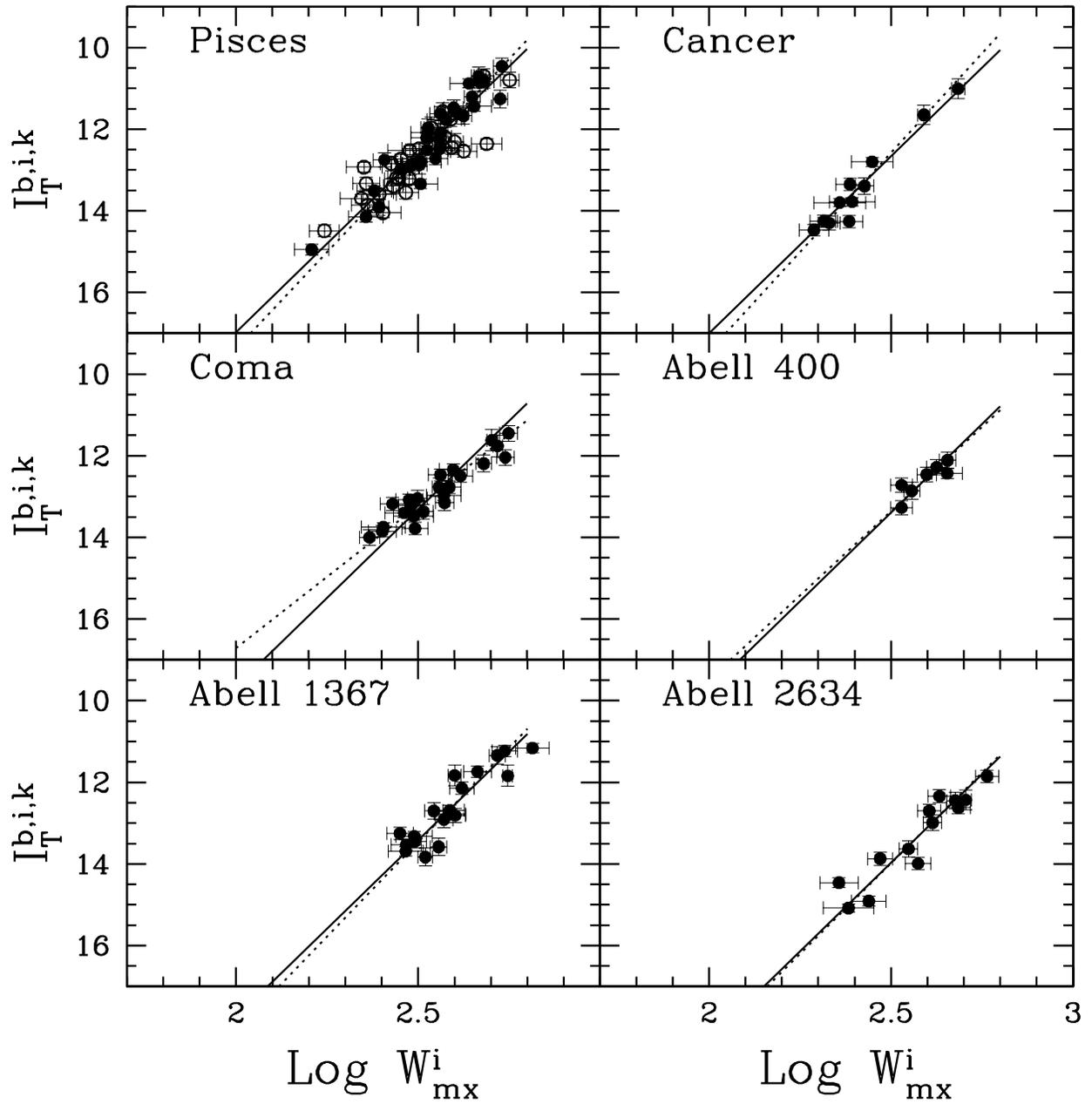}
\caption{$I$ band luminosity$-$linewidth correlations for 58 galaxies in Pisces, 11 galaxies in Cancer, 23 galaxies in Coma, 7 galaxies in A400, 19 galaxies in A1367, and 13 galaxies in A2634. Solid lines are the universal template fits and dotted lines are fits to the individual clusters. In the case of the Pisces filament galaxies lying at SGB$<5$ are identified by filled symbols and galaxies north of this limit are identified by open symbols.  See Appendix for discussions concerning Pisces and the other clusters.}
\label{tf6b}
\end{figure}

\twocolumn

\begin{deluxetable}{lrrrrrrrrrrrrr}
\tablenum{1}
\tablecaption{Properties of the Cluster Fits}
\label{tbl:clprop}
\tablewidth{0in}
\tablehead{\colhead{Cluster} & \colhead{$V_{CMB}$} & \colhead{No.} & \colhead{Slope} & \colhead{eSl} & \colhead{ZP} & \colhead{RMS} & b & \colhead{DM} & \colhead{eDM} & \colhead{Dist} & \colhead{eD} & \colhead{$V/$D} & \colhead{eH}}
\startdata
Virgo   &   1410 &   26 &   -8.83 &  0.52 &    9.62 &  0.44 &  0.00 & 31.01 &  0.11 &   15.9 &  0.8 &   88.7 &    5.5 \\
Fornax  &   1286 &   15 &   -8.57 &  0.56 &    9.80 &  0.41 & 0.00 & 31.19 &  0.12 &   17.3 &  1.0 &   74.3 &    5.2 \\
U Ma    &   1101 &   34 &   -8.51 &  0.42 &    9.81 &  0.49 &  0.00 & 31.20 &  0.10 &   17.4 &  0.9 &   63.3 &    4.4 \\
Antlia  &   3119 &   14 &   -9.83 &  1.03 &   11.42 &  0.30 &  0.04 & 32.85 &  0.10 &   37.2 &  1.7 &   83.8 &    4.2 \\
Cen30   &   3679 &   11 &  -10.66 &  1.53 &   11.51 &  0.52 & 0.01 & 32.91 &  0.17 &   38.2 &  3.1 &   96.3 &    8.0 \\
Pegasus &   3518 &   17 &   -7.56 &  0.76 &   11.89 &  0.42 &  0.00 & 33.18 &  0.12 &   43.4 &  2.5 &   80.9 &    4.8 \\
Hydra   &   4121 &   19 &   -8.57 &  0.94 &   12.42 &  0.54 &  0.03 & 33.84 &  0.14 &   58.6 &  3.8 &   70.3 &    4.8 \\
Pisces  &   4779 &   58 &   -9.45 &  0.56 &   12.62 &  0.40 &  0.01 & 34.02 &  0.08 &   63.7 &  2.4 &   75.0 &    3.0 \\
Cancer  &   4940 &   11 &   -9.73 &  0.74 &   12.65 &  0.26 &  0.02 & 34.06 &  0.10 &   64.9 &  3.0 &   76.1 &    3.7 \\
Coma    &   7194 &   23 &   -6.96 &  0.56 &   13.33 &  0.39 &  0.05 & 34.77 &  0.10 &   90.0 &  4.3 &   79.9 &    4.1 \\
A400    &   7108 &    7 &   -8.25 &  2.14 &   13.40 &  0.24 &  0.07 & 34.86 &  0.11 &   93.8 &  4.8 &   75.8 &    4.2 \\
A1367   &   6923 &   19 &   -9.26 &  1.03 &   13.42 &  0.39 &  0.05 & 34.86 &  0.11 &   93.8 &  4.7 &   73.8 &    3.9 \\
A2634   &   9063 &   13 &   -8.86 &  1.00 &   13.98 &  0.39 &  0.04 & 35.41 &  0.13 &  120.8 &  7.1 &   75.0 &    4.6 \\
\enddata

Column information:

\scriptsize{(1) Name; (2) velocity in CMB frame (\kms); (3) sample size; (4) slope determined for individual cluster; (5) slope uncertainty (68\%); (6) zero point for individual cluster assuming universal slope (mag); (7) r.m.s. scatter in magnitude for individual cluster; (8) bias correction (mag); (9) distance modulus determined for individual cluster (mag); (10) uncertainty in cluster distance modulus (mag); (11) distance determined for cluster (Mpc); (12) uncertainty in cluster distance (Mpc); (13) `Hubble parameter' (velocity/distance) for individual cluster (\kmsMpc); (14) uncertainty in Hubble parameter for individual cluster (\kmsMpc).}
\end{deluxetable}

\noindent
Here the dotted lines illustrate the individual fits and the solid lines give the common slope that is ultimately derived; the latter fit to the individual clusters with only the zero point as a free parameter.  Descriptors of the various fits are collected in Table~\ref{tbl:clprop}.  Clusters at successively greater distances have sample
cutoffs at successively brighter intrinsic luminosities.  However, on the
assumption that each sample is being drawn from the same universal relationship,
{\it the slope of the inverse relation is the same} (though less stable if
sampled over a shorter range).  By contrast, given the same scatter, the slope
of a direct or bivariate fit becomes shallower as the sampling range
diminishes!  The run of the slopes determined for individual clusters as a function of distance is shown in Figure~\ref{slopes}.
No dependence with distance is seen; there is no evidence here of bias.  The slopes for individual clusters do not differ significantly from the universal slope with the marginal exception of one case (Coma; see discussion in Appendix).

The nearest three clusters $-$ Virgo, Fornax, and Ursa Major $-$ are at
similar distances and are sampled over a similar range in magnitudes.  The
search for a universal relation begins by establishing the best common least squares fit slope and
relative zero point offsets that accommodate the three nearby clusters.  The offsets
were taken with respect to the Virgo Cluster and the fitting process quickly
converged.

The remaining ten clusters separate into three distance groups.  Antlia,
Centaurus, and Pegasus are $\sim 2$ mag more distant than Virgo; Hydra,
Pisces\footnote{The structure in Pisces is a filament rather than a cluster.  The filament lies very close to the plane of the sky with little or no variation in distance along its length even though it extends across 20 Mpc.  See discussion in Appendix.}, and Cancer are $\sim 3$ mag more distant; Coma, A1367, A400, and A2634
are $\sim 4$ mag more distant.  As a procedural matter, the template was
successively augmented with the addition of each of these three groups from
the nearer to the farther.  Optimal slopes and zero point offsets were
redetermined at each step.  Convergence was always rapid.  The result is a definition
of the universal slope and relative zero point offsets representing the
distance differences of each cluster with respect to the Virgo Cluster.

Interestingly, the scatters tend to decrease for the more distant clusters.
One reason can be that scatter increases toward lower intrinsic luminosities,
which are only sampled in the nearer clusters \citep{1997AJ....113...22G}.    It has been demonstrated \citep{2000ApJ...533L..99M, 2005ApJ...632..859M}
that scatter at faint luminosities can be reduced with a recipe that accounts
for the increasing importance of interstellar gas to the inventory of baryonic
matter.  Another reason can be that the nearby clusters are being resolved;
the assumption that the targets are at a common distance is breaking
down.  Some attention to this matter is given in the Appendix for individual clusters.

The cummulative plot with clusters shifted to optimally fit the relationship
at the Virgo distance is shown in Figure~\ref{13atvirgo}.  This `template' of the new
$I$~band $-$ HI linewidth correlation is built with 267 galaxies in 13
clusters extending in range from 1000 \kms\ to 10,000 \kms.

\renewcommand{\thefigure}{\arabic{figure}}
\begin{figure}[t]
\includegraphics[scale=0.39]{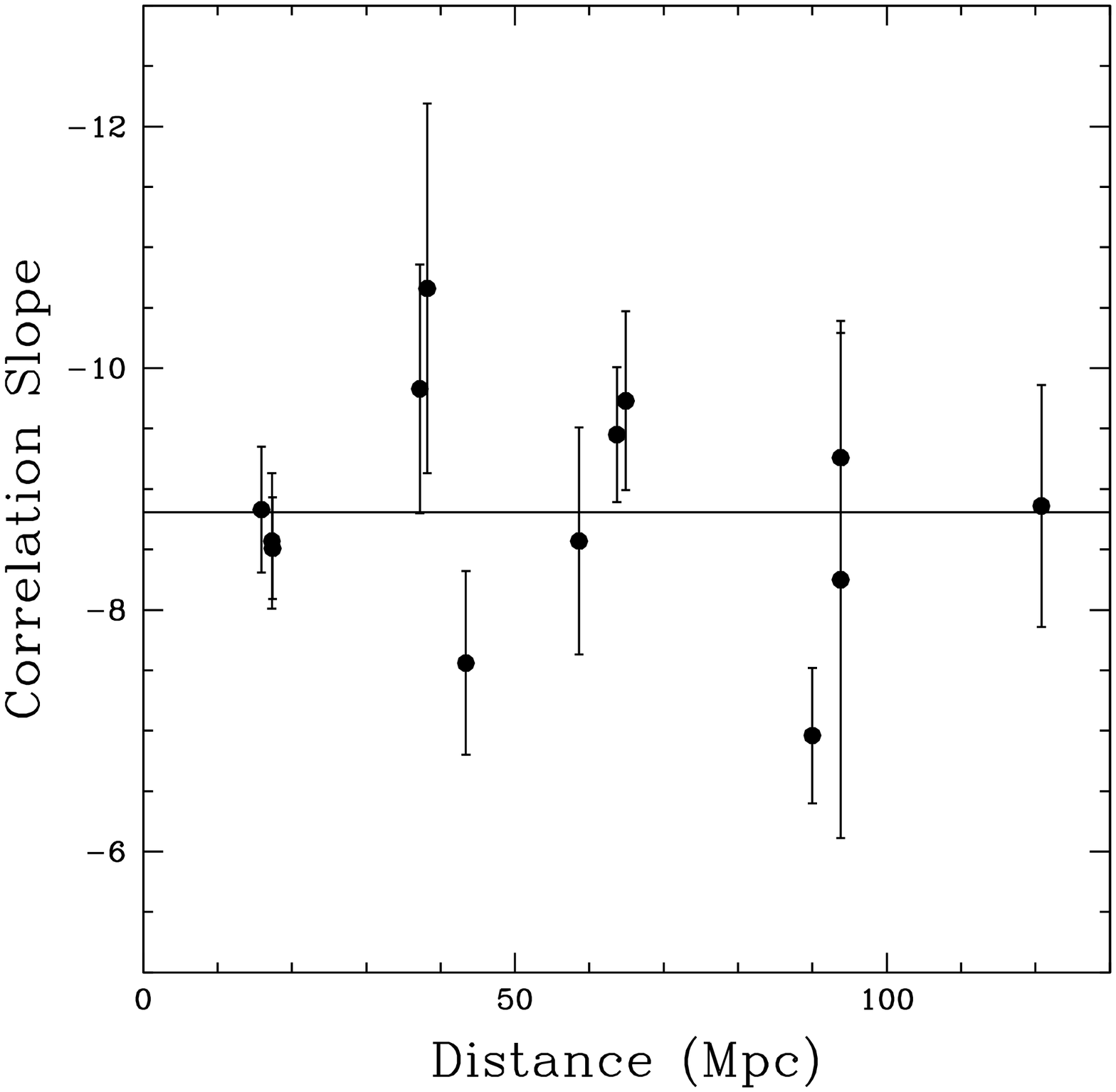}
\caption{Correlation slope determined for individual clusters as a function of distance.  The solid line shows the slope of $-8.81$ given by the fit to the ensemble of 13 clusters.}
\label{slopes}
\end{figure}

\begin{figure}[t]
\includegraphics[scale=0.39]{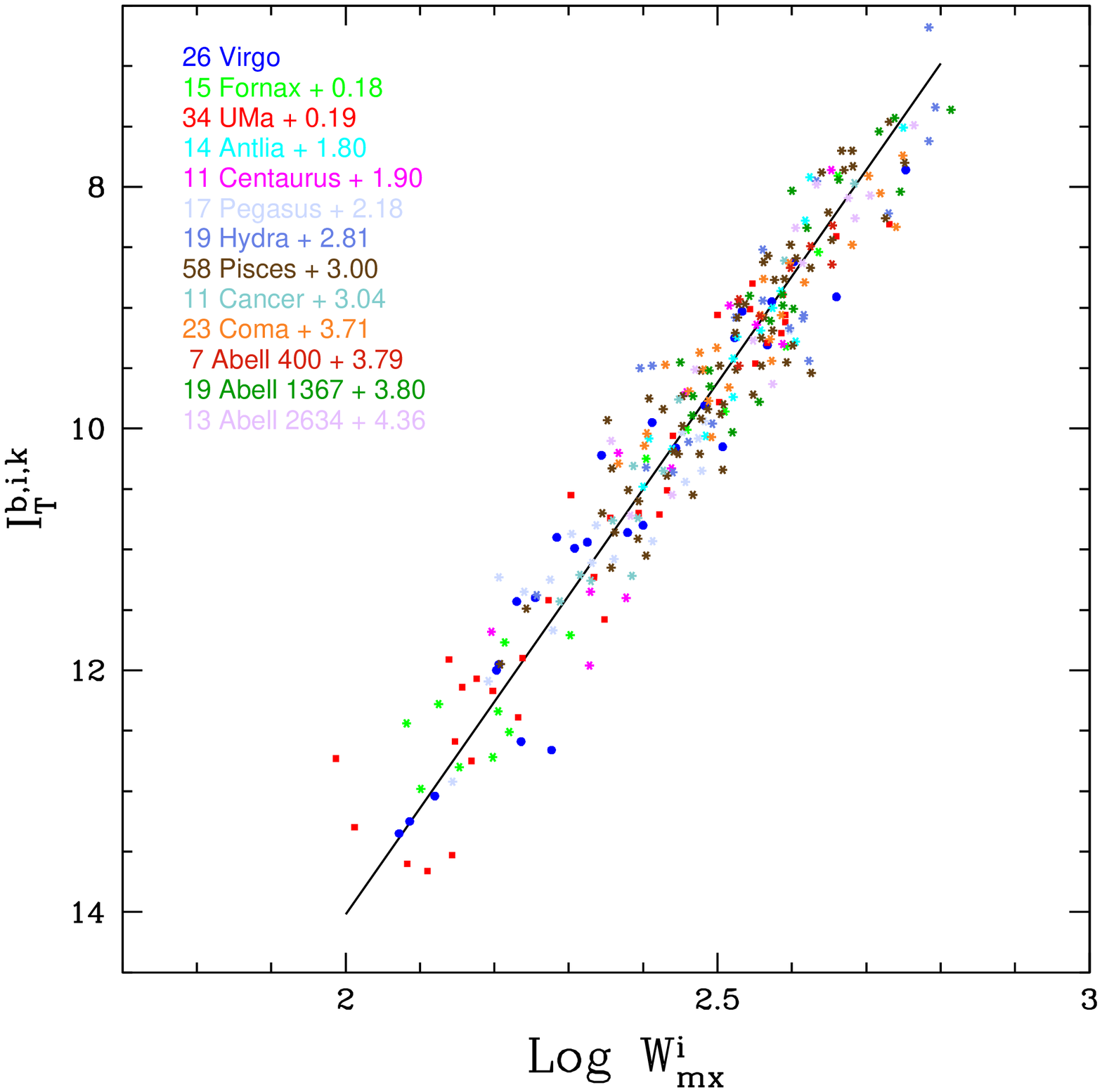}
\caption{Cumulative plot superimposing data from 13 clusters with zero point
  offsets relative to the Virgo Cluster.  Solid line is least squares fit to
  the ensemble with errors in linewidths.}
\label{13atvirgo}
\end{figure}

\begin{deluxetable}{rlrcrrccl}
\tablenum{2}
\tablecaption{Zero Point Calibrators}
\label{tbl:zp}
\tablewidth{0in}
\tablehead{\colhead{PGC} & \colhead{Name} & \colhead{$I_T^{b,i,k}$} & \colhead{Inc} & \colhead{$W_{mx}$} & \colhead{$W_{mx}^i$} & \colhead{log $W_{mx}^i$} & DM & \colhead{$M_I^{b.i.k}$}}
\startdata
   1014 & NGC0055  &   6.88 &   86 &    166 &    166 &   2.221 &  26.62 &  -19.74 \\
   2557 & NGC0224  &   1.34 &   78 &    517 &    528 &   2.722 &  24.48 &  -23.14 \\
   2758 & NGC0247  &   7.79 &   76 &    190 &    196 &   2.292 &  27.84 &  -20.05 \\
   2789 & NGC0253  &   5.07 &   81 &    410 &    415 &   2.618 &  27.83 &  -22.76 \\
   3238 & NGC0300  &   7.28 &   46 &    140 &    195 &   2.290 &  26.59 &  -19.31 \\
   5818 & NGC0598  &   4.74 &   55 &    177 &    217 &   2.336 &  24.82 &  -20.08 \\
   9332 & NGC0925  &   8.96 &   57 &    194 &    231 &   2.364 &  29.81 &  -20.85 \\
  13179 & NGC1365  &   8.09 &   54 &    371 &    459 &   2.662 &  31.27 &  -23.18 \\
  13602 & NGC1425  &   9.50 &   65 &    354 &    391 &   2.592 &  31.70 &  -22.20 \\
  17819 & NGC2090  &   9.33 &   67 &    277 &    301 &   2.478 &  30.35 &  -21.02 \\
  21102 & NGC2366  &  11.03 &   74 &     90 &     94 &   1.972 &  27.57 &  -16.54$\star$ \\
  21396 & NGC2403  &   7.11 &   60 &    226 &    261 &   2.417 &  27.50 &  -20.39 \\
  23110 & NGC2541  &  10.76 &   63 &    188 &    211 &   2.325 &  30.25 &  -19.49 \\
  26512 & NGC2841  &   7.53 &   66 &    592 &    650 &   2.813 &  30.74 &  -23.21 \\
  28120 & NGC2976  &   8.98 &   60 &    129 &    149 &   2.173 &  27.77 &  -18.79 \\
  28357 & NGC3021  &  10.92 &   57 &    254 &    303 &   2.481 &  32.27 &  -21.35 \\
  28630 & NGC3031  &   5.20 &   59 &    417 &    485 &   2.686 &  27.81 &  -22.61 \\
  29128 & NGC3109  &   9.15 &   90 &    110 &    110 &   2.041 &  25.62 &  -16.47$\star$ \\
  30197 & NGC3198  &   9.17 &   70 &    296 &    315 &   2.498 &  30.70 &  -21.53 \\
  30819 & IC2574   &  10.12 &   69 &    106 &    113 &   2.054 &  27.96 &  -17.84 \\
  31671 & NGC3319  &  10.55 &   59 &    195 &    227 &   2.356 &  30.62 &  -20.07 \\
  32007 & NGC3351  &   8.33 &   47 &    262 &    359 &   2.556 &  30.00 &  -21.67 \\
  32207 & NGC3370  &  10.85 &   58 &    264 &    312 &   2.494 &  32.13 &  -21.28 \\
  34554 & NGC3368  &   8.01 &   66 &    266 &    292 &   2.465 &  29.08 &  -21.07 \\
  34695 & NGC3621  &   7.39 &   60 &    333 &    385 &   2.585 &  29.59 &  -22.20 \\
  39422 & NGC4244  &   8.92 &   90 &    192 &    192 &   2.283 &  28.16 &  -19.24 \\
  39600 & NGC4258  &   6.84 &   69 &    415 &    444 &   2.647 &  29.41 &  -22.57 \\
  40692 & NGC4414  &   8.73 &   55 &    378 &    463 &   2.666 &  31.24 &  -22.51 \\
  41812 & NGC4535  &   8.95 &   45 &    265 &    374 &   2.573 &  30.99 &  -22.04 \\
  41823 & NGC4536  &   9.03 &   71 &    322 &    341 &   2.533 &  30.91 &  -21.88 \\
  42408 & NGC4605  &   9.19 &   69 &    154 &    165 &   2.219 &  28.71 &  -19.52 \\
  42510 & NGC4603  &   9.76 &   52 &    353 &    450 &   2.653 &  32.60 &  -22.84 \\
  42741 & NGC4639  &  10.18 &   55 &    274 &    336 &   2.526 &  31.67 &  -21.49 \\
  43451 & NGC4725  &   7.84 &   58 &    397 &    470 &   2.672 &  30.53 &  -22.69 \\
  45279 & NGC4945  &   6.30 &   86 &    358 &    359 &   2.555 &  27.63 &  -21.33 \\
  51344 & NGC5584  &  10.62 &   44 &    186 &    267 &   2.426 &  31.72 &  -21.10 \\
  69327 & NGC7331  &   7.52 &   66 &    501 &    547 &   2.738 &  30.72 &  -23.20 \\
  73049 & NGC7793  &   8.25 &   53 &    162 &    202 &   2.306 &  27.79 &  -19.54 \\
\enddata

Column information:

\scriptsize{(1) Principal Galaxy Catalog ID; (2) name; (3) apparent $I$ band magnitude corrected for Galactic and internal reddening and redshift; (4) inclination from face on; (5) linewidth at 50\% of mean flux with adjustments to statistically represent twice maximum projected rotation velocity (\kms); (6) linewidth parameter de-projected to edge on orientation (\kms); (7) logarithm of de-projected linewidth parameter; (8) distance modulus from independent (Cepheid or TRGB) distance measurement; (9) Absolute $I$ band magnitude at the independently determined distance. The two alerts signaled by $\star$ identify galaxies fainter than the range encompassed by the template slope calibration and not used in the zero point calibration.}
\end{deluxetable}

\section{Intrinsic Zero Point}
\label{sec:zp}

There are increasing numbers of relatively nearby galaxies appropriate for
the luminosity$-$linewidth methodology with independent distance measures of
reasonable accuracy.  We accept the zero point established by the Hubble
Space Telescope distance scale project  \citep{2001ApJ...553...47F} and the
distances to galaxies from Cepheid Period$-$Luminosity measures consistent
with that scale.  We also accept measures from the Tip of the Red Giant Branch
method that have been shown to be on a consistent scale \citep{2007ApJ...661..815R}.
The maser distance to NGC4258 is also taken into account \citep{1999Natur.400..539H}.
Information about the zero point calibration sample is collected in Table~\ref{tbl:zp}.

\begin{figure}[h]
\includegraphics[scale=0.39]{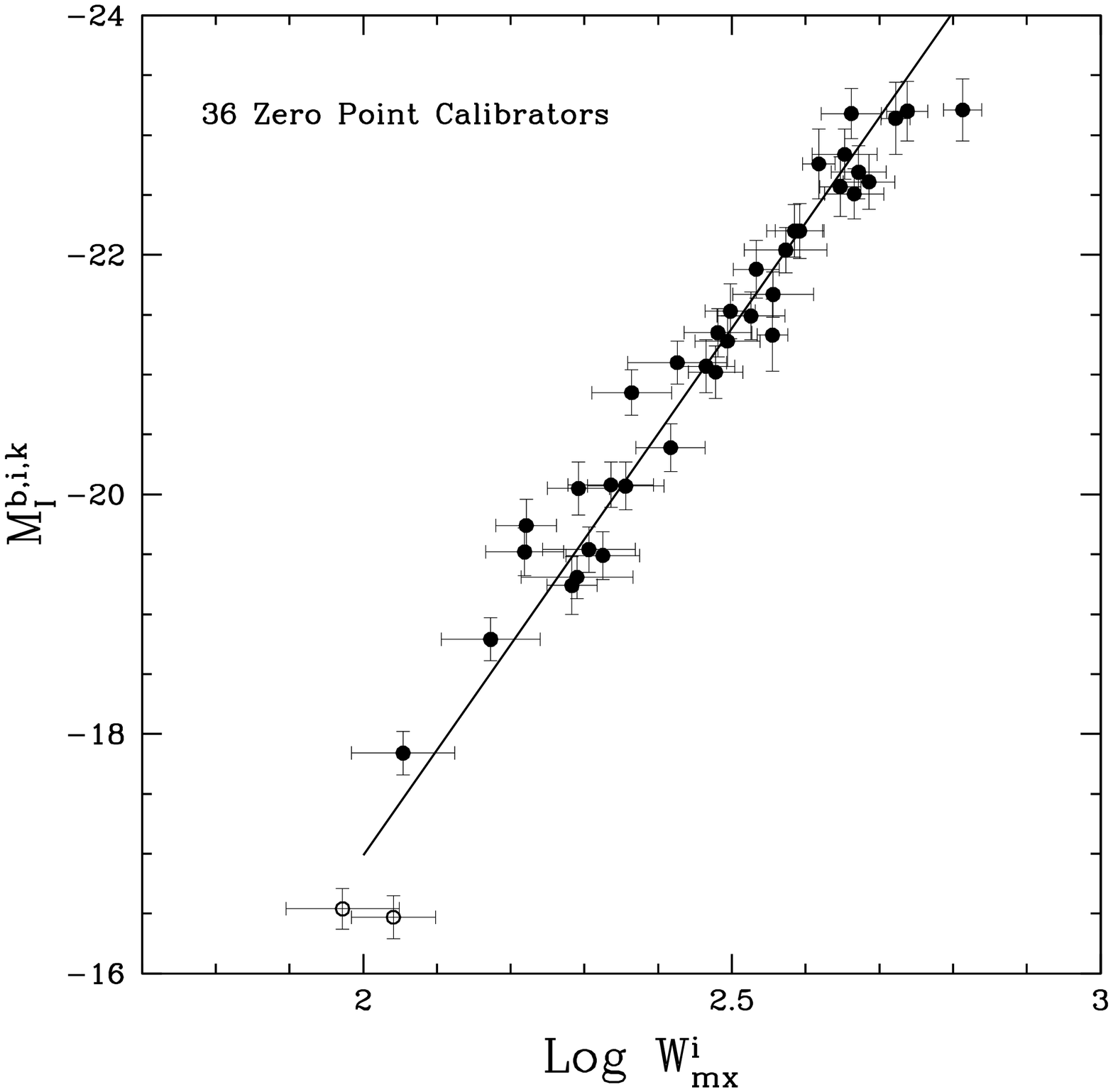}
\caption{Zero point calibration based on 36 galaxies with distances
  independently established with either Cepheid Period$-$Luminosity or Tip of
  the Red Giant Branch measures.  The solid line is the least squares fit with
the slope fixed to the value established by the 13 cluster template.}
\label{tfzp}
\end{figure}

The luminosity$-$linewidth relation found with  these galaxies with
independent distances is shown in Figure~\ref{tfzp} now on an intrinsic luminosity scale.  These galaxies do not represent
a magnitude limited sample so a slope fit to these data would be inappropriate.  The
fit illustrated by the solid line has the slope defined by the 13 cluster
template and a zero point established from a least squares fit to the 36 calibrators that lie
within the magnitude range of the template.  The sample of zero  point
calibrators is consistent with being drawn from the universal correlation.

\begin{figure}[t]
\includegraphics[scale=0.39]{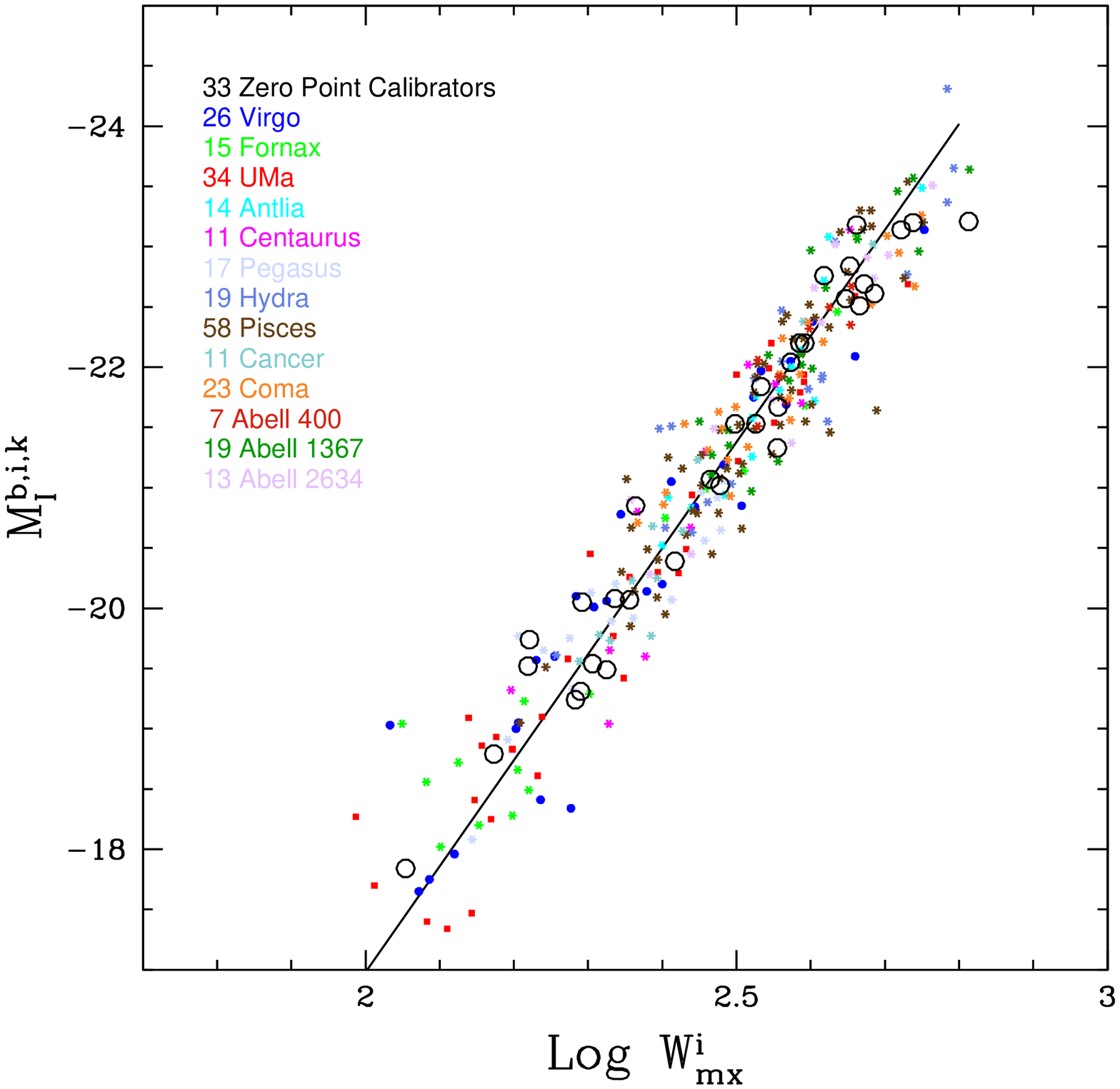}
\caption{The zero point calibrator and 13 cluster template samples combined.
  The absolute magnitude scale for the cluster sample is set by the fit to the
zero point calibrators.}
\label{13atzp}
\end{figure}

The magnitude scale relations between the fits in Figures~\ref{13atvirgo} and \ref{tfzp} give a distance
modulus for the Virgo Cluster and, from the offsets built into Figure~\ref{13atvirgo}, give
distance moduli for the other clusters.  The combined cluster and zero point calibrator samples are shown together in Figure~\ref{13atzp}.  The associated distances are
recorded in Table~\ref{tbl:clprop}.  If our fully corrected luminosity and HI profile parameters are employed then the distance modulus of a galaxy $\mu = I^{b,i,k} - M_I^{b,i,k}$ can be calculated from the relation between linewidth and absolute magnitude
\begin{equation}
M_I^{b,i,k} = -21.39 -8.81 ({\rm log} W^i_{mx} - 2.5)
\label{absM}
\end{equation}
where $1\sigma$ uncertainties are $\pm0.07$ in the zero point and $\pm0.16$ in the slope.  The r.ms. scatter about the mean relation for the ensemble of 267 template galaxies is $\pm0.41$ mag corresponding to a scatter in distance of 21\%.  Some of this scatter must be contributed by the finite depth of the clusters.  The r.m.s. scatter about the mean for the 36 zero point calibrators with Cepheid or TRGB distances is $\pm0.36$, not substantially less.

\begin{figure}[t]
\includegraphics[scale=0.39]{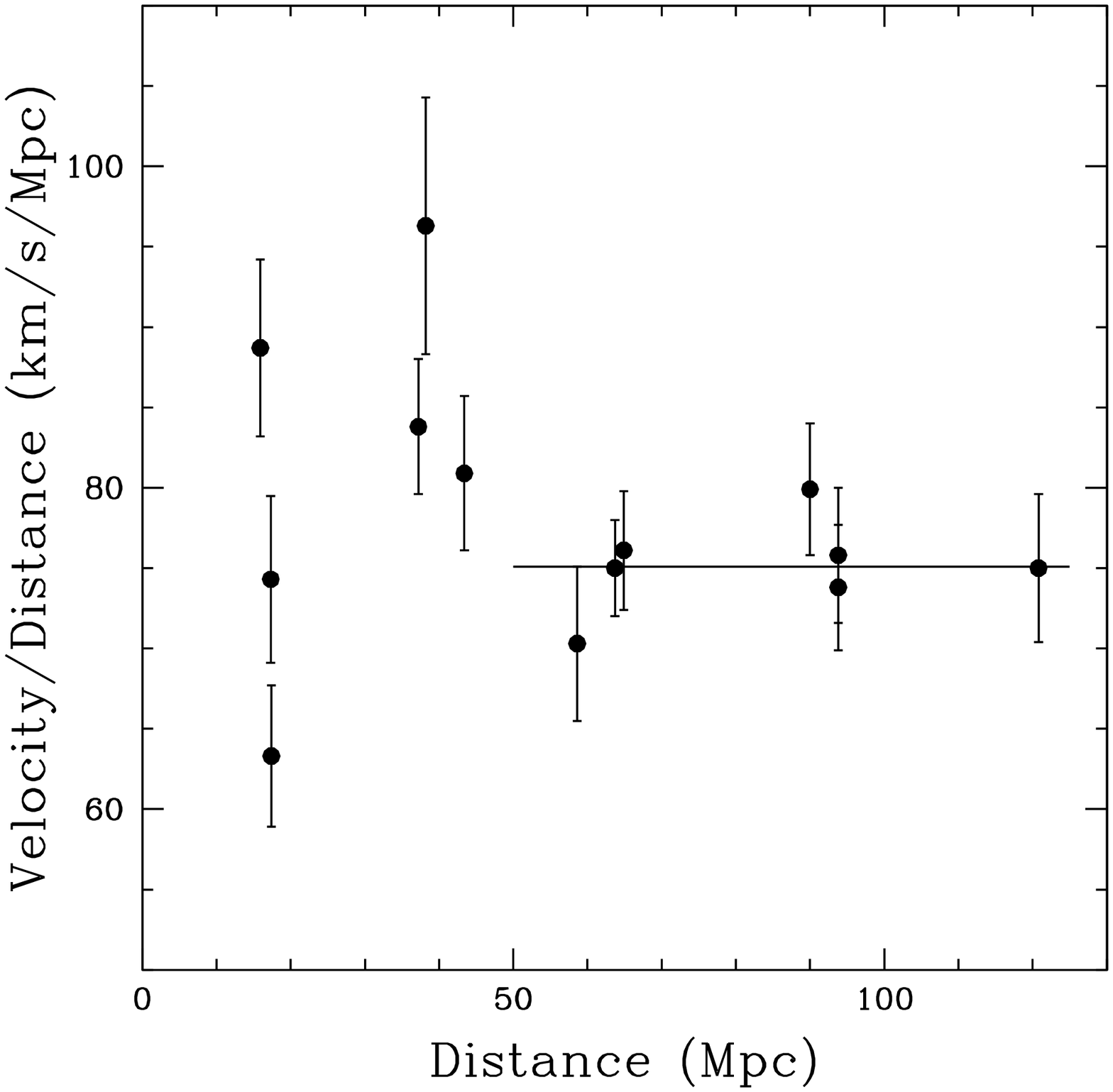}
\caption{Run of the Hubble parameter (velocity in the CMB frame / distance) as a function of distance.  Fit to cluster points at distances greater than 50 Mpc ($V_{CMB} > 4000$~\kms) corresponds to H$_0 = 75.1$ \kmsMpc.}
\label{dH}
\end{figure}

Hubble parameters (velocity/distance) for the 13 clusters are displayed as
a function of velocity in the cosmic microwave background frame in Figure~\ref{dH}.  The Hubble parameter gyrates wildly within 50 Mpc ($V_{CMB} < 4000$~\kms), the region of the Local Supercluster and the Great Attractor, but shows little variation for the 7 most distant clusters.  A fit to those outer 7 clusters gives the result H$_0=75.1\pm2.7$~\kmsMpc\ with a standard deviation of $\pm1.0$~\kmsMpc.

\section{Literature Comparisons}

We first make a comparison with our earlier calibrations using similar procedures.   The new calibration uses digital rather than analog HI information with a different definition of linewidth and enlarged slope and zero point samples but distances for 12 clusters in common with the  \citet{2000ApJ...533..744T} study are overall only slightly increased (new cluster distances are larger by 1.4\%) although the most distant clusters are set back 5\%, mostly due to the bias correction that is now introduced.  Tully \& Pierce found H$_0 = 77$~\kmsMpc.  Studies with the same methodology a decade earlier \citep{1992ApJ...391...16S} had suggested H$_0 = 88$~\kmsMpc\ but Tully \& Pierce pointed out that most of the reduction in H$_0$ by 2000 resulted from an improved zero point with the availability of Cepheid distances from the HST distance scale key project. 

Now giving attention to studies from independent researchers, we find that
there are sufficient overlaps for meaningful comparisons with two existing samples.  The first one to consider is a separate analysis with the same methodology.  We give attention to the Cornell group results for clusters presented by \citet{2006ApJ...653..861M}.  Before making the comparison we undid a correction made by Masters et al. that appears to us to be erroneous. They correctly point out that the true distance of a cluster is statistically greater than the value implied by the average of the measured moduli but they overestimate the effect by an order of magnitude.  They give `corrections' for the effect that range from 5\% for the nearest clusters to 0.5\% for the most distant.  From simulations we determine corrections of 0.5\% for the nearest 3 clusters (which we incorporate) and negligible corrections for the rest.  As a practical matter, only Fornax and Ursa Major are affected in the present comparison since Masters et al. did not consider the Virgo Cluster.

Our separate results are given in Table~\ref{tbl:compD} for the 12 clusters in common (ie, excluding Virgo).  There is good agreement.  It is to be recognized that the results are hardly independent.  Most (not all) of both the raw HI profile and $I$-band photometry data are used in common.  Both studies avail of Cepheid distances from the HST Key Project to set zero points.  Looking in detail, the worst individual match is in the case of the Pegasus Cluster where the disagreement is at the level of $2\sigma$.  There is the disturbing systematic of poor agreement with the clusters nearer than 4000~\kms.  Masters et al. place these 5 clusters more distant by 10\% with a statistical significance of $4\sigma$.  On the other hand there is agreement with the 7 more distant clusters at the level of 2\% in the mean and with a per cluster scatter of only 2\%.

The disagreement with the nearer clusters remains to be clarified.  If attention is restricted to clusters beyond 4000~\kms, it is seen from Figure~\ref{dH} that we derive H$_0 = 75$~\kmsMpc\ from this limited sample.  Masters et al. determined a value of H$_0=74$~\kmsMpc\ from a more extensive sample beyond 4000~\kms. 

The second comparison to be made is with Fundamental Plane measurements where the material is entirely independent.  We compare two distinct Fundamental Plane samples: one with the acronym ENEARc \citep{2002AJ....123.2990B} and the other called SMAC \citep{2001MNRAS.327..265H}.  We mention in passing that we have re-analyzed both these data sets using `inverse' relationships analogous to the `inverse' relation between rotation and luminosity in spirals and we find distances in excellent agreement with those found by the authors of the ENEARc and SMAC studies; in other words, the inverse fits effectively null bias without corrections.  However for the present comparisons we draw on the distance determinations given by the original authors.

The zero point that we accept for the Fundamental Plane measurements comes from the comparison between this method with the SMAC sample and the Surface Brightness Fluctuation method \citep{2002MNRAS.330..443B}.  Tiny corrections are made to the Surface Brightness Fluctuation distances following \citet{2010ApJ...724..657B}.  The zero point is extended to the ENEARc sample by a comparison of distances for 18 clusters in common to the ENEARc and SMAC studies.

\begin{figure}[t]
\includegraphics[scale=0.39]{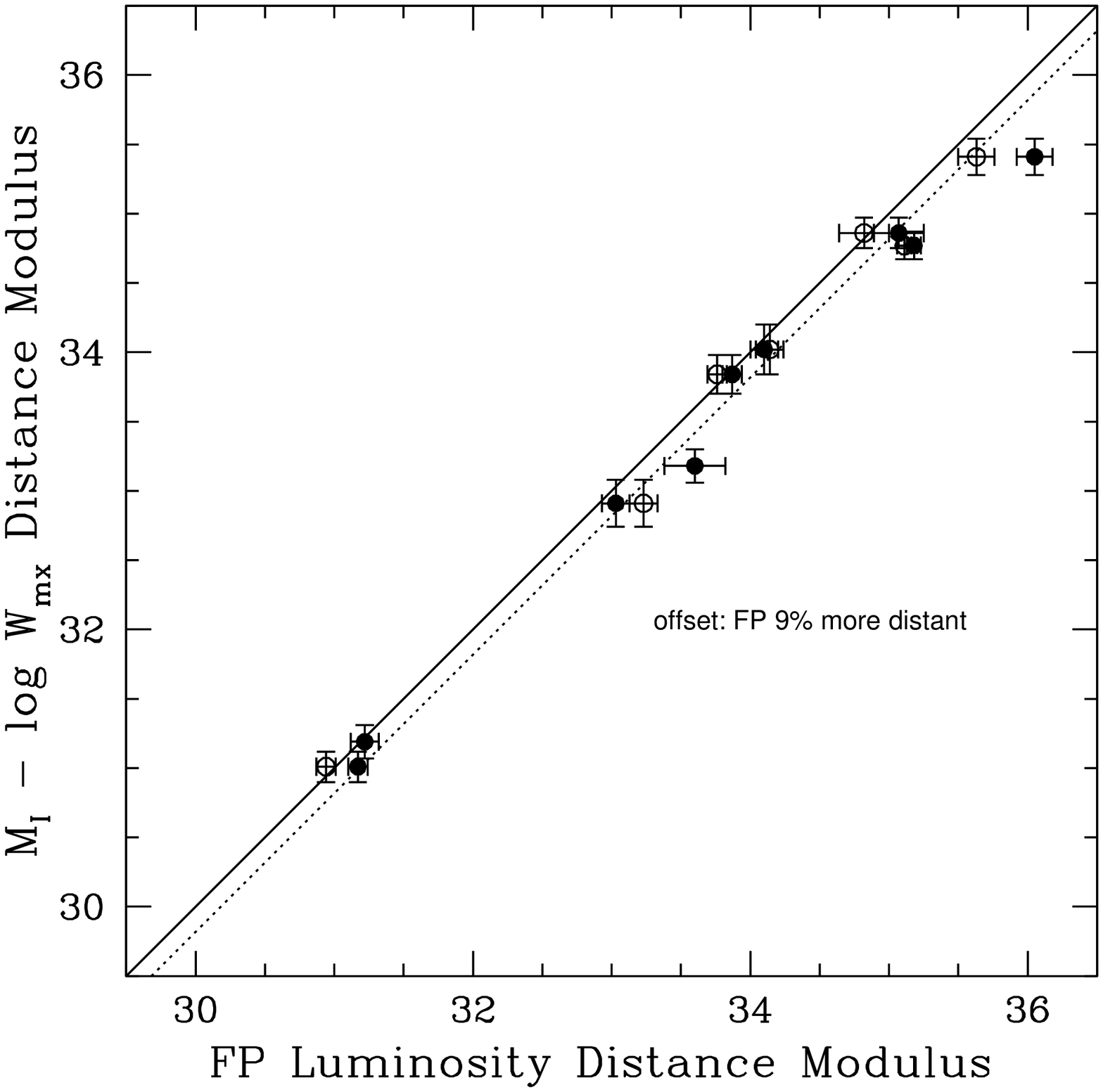}
\caption{Comparison of distance moduli from this paper with Fundamental Plane moduli.  Filled symbols: ENEARc; open symbols: SMAC.}
\label{fp}
\end{figure}

Alternative distances for clusters in our program are reported in Table~\ref{tbl:compD} and graphical comparisons are made in Figure~\ref{fp}.  A systematic offset is seen.  If we accept a slope of unity in the fit to measures as a function of distance then the offset is 9\% in distance, Fundamental Plane measures larger, with $4\sigma$ significance and 10\% scatter per individual cluster.  There is a hint that the slope in the data of Figure~\ref{fp} is slightly flatter than unity but the significance is only $2\sigma$.

There are some issues that remain to be reconciled.  With the Surface Brightness Fluctuation zero point, the cumulative ENEARc and SMAC samples are consistent with H$_0=72$~\kmsMpc.  A renormalization of those samples to agree with our distances for 9 of their clusters would imply H$_0=78$~\kmsMpc.  On the other hand, \citet{2008ApJ...676..184T} report agreement between Surface Brightness Fluctuation and luminosity-linewidth distances in Cosmicflows-1.  These conflicts will not be resolved in the present work.  They will be given further attention with the integration of distances derived with these methods and several others during the compilation of Cosmicflows-2.

\begin{deluxetable}{lccccc}
\tablenum{3}
\tablecaption{Comparison with Literature Distances}
\label{tbl:compD}
\tablewidth{0in}
\tablehead{\colhead{Cluster} & \colhead{This Paper} & \colhead{TP00} & \colhead{SFI++} & \colhead{ENEARc} & \colhead{SMAC}}
\startdata
Virgo    & $15.9\pm0.8$ & $   -     $ &   $ -$                 & $17.1\pm0.8$ & $15.4\pm0.9$ \\
Fornax & $17.3\pm1.0$ & $ 17.8 $ & $19.6\pm1.3$ & $17.5\pm1.1$ &   $ -$                 \\
U Ma    & $17.4\pm0.9$ & $ 18.6 $ & $17.9\pm1.2$ &   $ -$                 &   $ -$                 \\
Antlia   & $37\pm2$        & $ 36    $ & $40\pm2$        &    $-$                 &    $-$                 \\
Cen30  & $38\pm3$       & $ 40    $ & $40\pm2$        & $40\pm2$        &  $44\pm2$       \\
Pegasus & $43\pm3$     & $ 46    $ & $51\pm3$        & $52\pm6$        &  $-$                   \\
Hydra   & $59\pm4$        & $ 58    $ & $58\pm3$        & $59\pm3$        &  $56\pm3$       \\
Pisces  & $64\pm2$       & $ 60    $ & $64\pm4$         & $66\pm4$        &  $67\pm3$       \\
Cancer & $65\pm3$       & $ 62    $ & $65\pm4$         & $-$                    &  $-$                   \\
Coma   & $90\pm4$       & $ 86    $ & $91\pm5$         & $109\pm5$      &  $105\pm4$     \\
A400    & $94\pm5$       & $ 92    $ & $91\pm5$         & $-$                     &  $-$                    \\
A1367  & $94\pm5$       & $ 86    $ & $90\pm5$         & $103\pm10$    &  $92\pm7$       \\
A2634  & $121\pm7$     & $ 111  $ & $116\pm7$       & $162\pm12$    &  $134\pm6$    \\
\enddata

Notes:

\scriptsize{Distances and errors in Mpc.  Literature sources: TP00 \citep{2000ApJ...533..744T}; SFI++  \citep{2006ApJ...653..861M}; ENEARc \citep{2002AJ....123.2990B}; SMAC  \citep{2001MNRAS.327..265H}. }
\end{deluxetable}

\section{Conclusions}

A re-calibration of the relationship between galaxy rotation rates and luminosities is necessitated by the initiation of a new way of measuring HI profile linewidths.  Our purpose is to measure distances and in this case we advocate use of the `inverse' relationship that almost nulls the magnitude limit Malmquist (selection) bias.  There is a residual correction that affects the measure of H$_0$ at the level of 2\%.  If the purpose were to try to understand the origins of the luminosity-rotation relationship it would be advisable to give attention to a description given by the bivariate fit.  In this case and with our parameters, $L_I \propto W^{3.4\pm0.1}$.

There can be concern about whether the relationship is the same in different environments; ie, whether there is a `universal' form at the present epoch.  We have explored the situations in 13 different environments ranging from high density collapsed regions dominated by early-type systems to extended low density regions dominated by gas-rich systems (see the Appendix for details).  The samples from all 13 environments are consistent with being drawn from a single universal distribution.  \citet{2006ApJ...653..861M} have reached a similar conclusion.  It is true that there are more aberrant cases in the collapsed cores of clusters.  The harsh environments of clusters probably do have an effect.  For our purposes there may be an ameliorating circumstance.  The spiral galaxies seen in and around rich clusters may mostly be recent arrivals.  Anyway, empirically there is no evidence contrary to the proposition that the calibration is representative.  

The scatter in magnitude about the mean relationship is characteristically 0.4 mag which translates to 20\% uncertainty in a distance estimate.  As \citet{1997AJ....113...22G} have pointed out, the scatter is less at high luminosities and greater at low luminosities.  Fractional uncertainties are largest for fainter galaxies with narrower linewidths and for more face on galaxies requiring larger de-projection corrections.  As \citet{2001ApJ...563..694V} has demonstrated, the scatter can be reduced with sample pruning and resolved kinematic information.  There are claims of type and surface brightness dependencies.  These diminish toward the infrared and with the steeper `inverse' slope to the extent that the second parameter dependencies become marginal and undesirable to implement because they introduce small discontinuities and subjectivity that may be distance dependent.  Sensitivity to distance diminishes with the steeper inverse slope but the offsetting benefits are diminished bias and enhanced simplicity.

The comparison with alternative distance measures, particularly Fundamental Plane measures, demonstrate overall agreement and small scatter but there are still unresolved zero point issues at the 9\% level.  The full construction of Cosmicflows-2 will involve information from considerably more sources and hopefully the zero point issues will be resolved or at least diminished.  From the fits with our data alone and restricted to the 7 clusters at $V_{CMB} > 4000$~\kms\ we infer H$_0=75.1$~\kmsMpc. The formal standard deviation is a ridiculously small 1~\kmsMpc\ but the sample is small, there are remaining zero point uncertainties, and then, as hinted at with the large local deviations in Figure~\ref{dH}, there is  the undiscussed issue of velocity perturbations on large scales.  There is still work to do.

\bigskip\bigskip\noindent
Individuals that have helped with the collection and analysis of contributing data include Austin Barnes, Nicolas Bonhomme, Rick Fisher, Philippe H\'eraudeau, Dmitry Makarov, Luca Rizzi, and Max Zavodny.  Material for the Fundamental Plane comparison that supplemented published information was supplied by John Blakeslee and Mike Hudson.  The component of our HI profile information that is new comes from observations in the course of the Cosmic Flows Large Program with the NRAO Green Bank Telescope augmented by observations with Arecibo and Parkes telescopes.  Support has been provided by the US National Science Foundation with award AST-0908846. 

\bigskip\noindent{\bf
{Appendix: Cluster Sample Details}}

Galaxy clusters are rarely simple structures.  The following discussion summarizes issues arising with each of the 13 clusters.

\begin{figure}[!]
\centering
\includegraphics[scale=0.38]{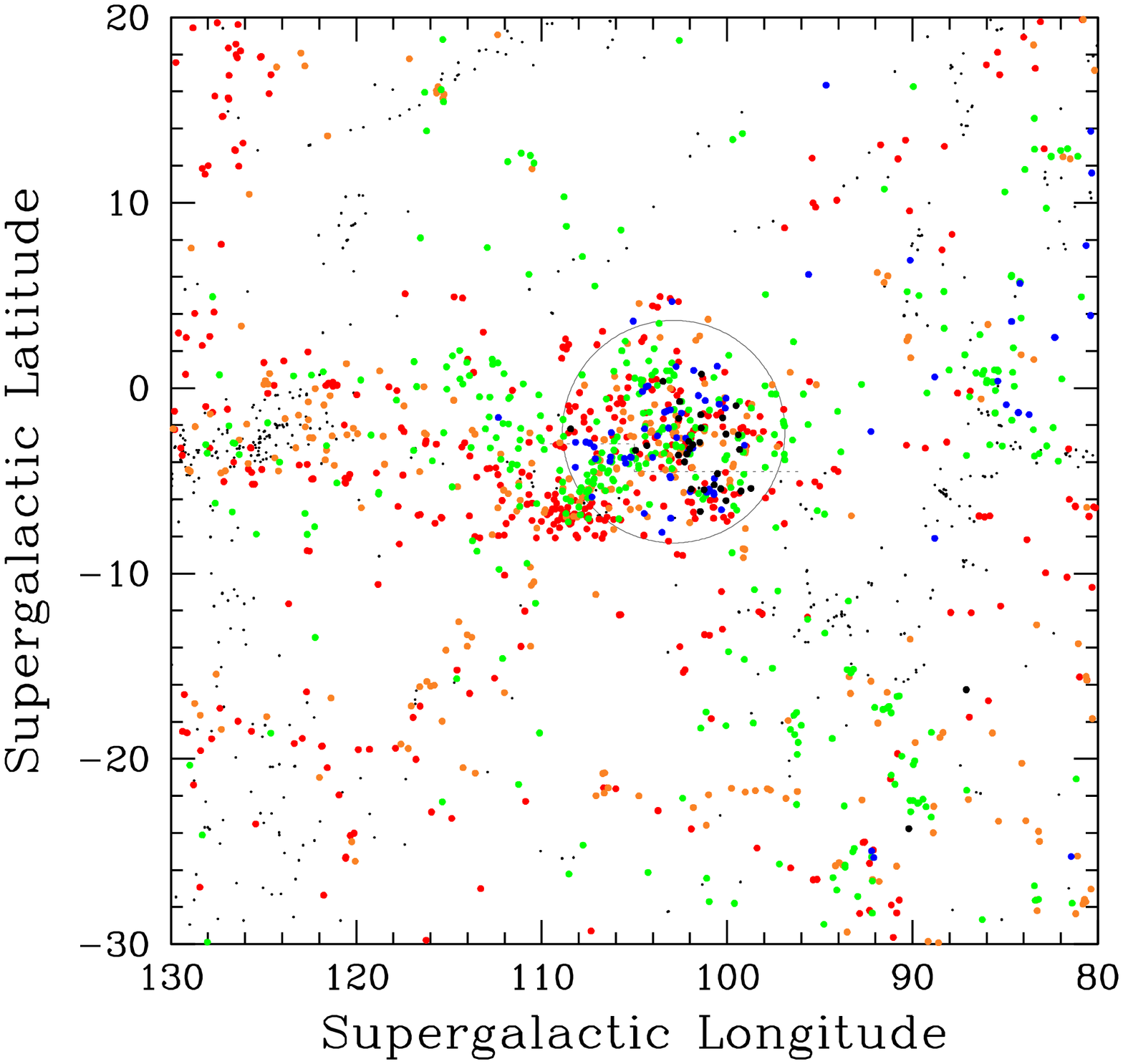}
\includegraphics[scale=0.3]{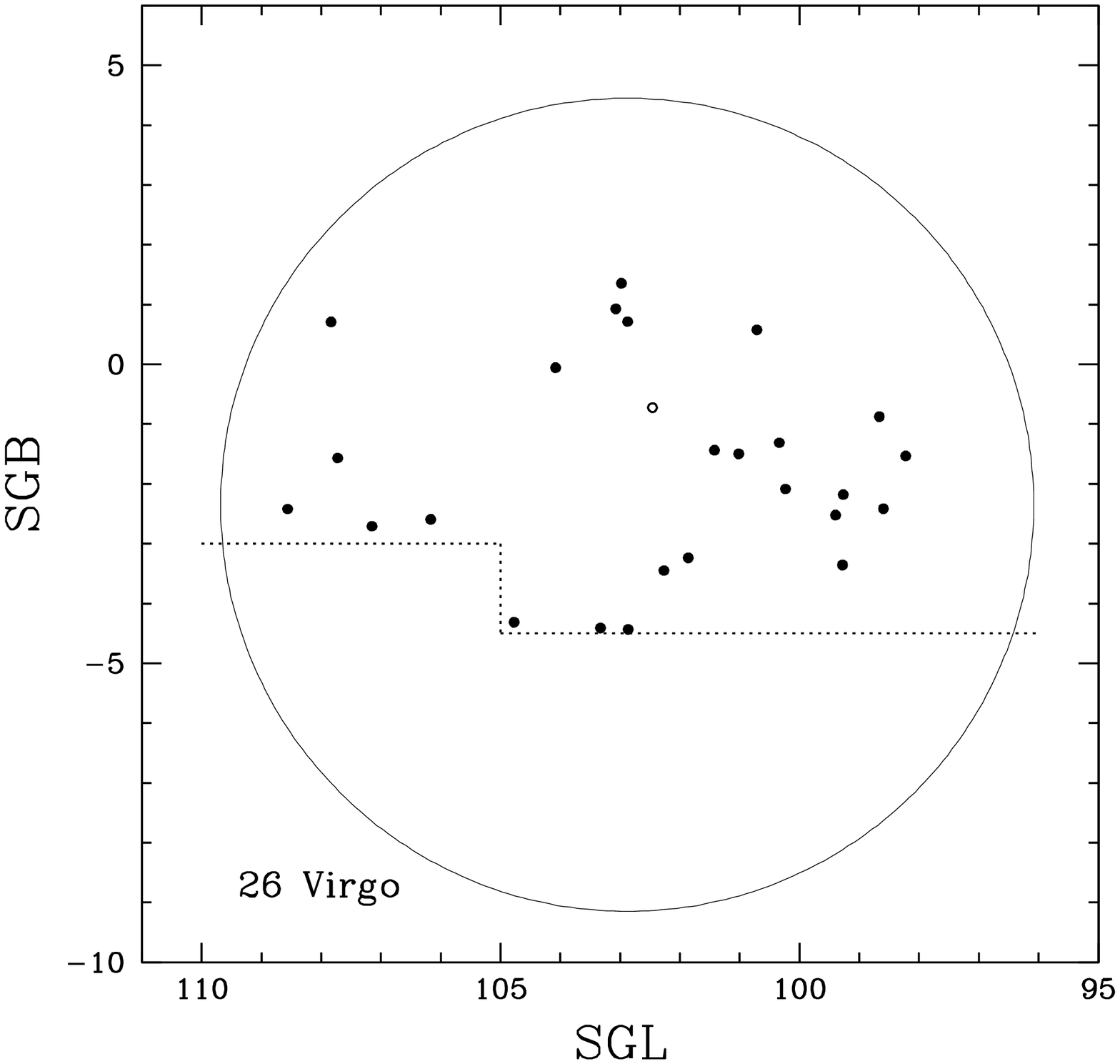}
\caption{{\it Top.} The distribution of galaxies near the Virgo Cluster.  Systemic velocity intervals are distinguished by colors: black (negative velocities), blue (0-500 \kms), green (500-1100), orange (1100-1500), red (1500-3000), small black dots (3000-6000).  The collapsed cluster lies within the circle at 2 Mpc radius.  Most infalling galaxies are arriving from the Southern Extension lying near the supergalactic equator at relatively positive supergalactic longitudes.  The background Virgo W cluster lies at SGL=109, SGB=$-$7, near the edge of the Virgo Cluster.  {\it Bottom.} The distribution of Virgo galaxies used in the current study is shown in this blow-up of the Virgo Cluster core.  Known background contaminants all lie below the broken dotted lines.  The open symbol locates the probable foreground galaxy PGC42081=IC3583.}
\label{lbvir}
\end{figure}

\noindent{\bf Virgo Cluster.}  This cluster was not included in the catalog by  \citet{1958ApJS....3..211A} because of its proximity but it comfortably qualifies as an Abell cluster of richness class zero.  It is a good candidate for our study because it contains many spirals as well as early type systems.  Probably many of the spirals in the cluster are recent arrivals because galaxies are seen to be falling into the cluster along a filament known as the Virgo Southern Extension \citep{1984ApJ...281...31T}.  Within the cluster, galaxies have observed velocities ranging between -600 and 2800 \kms.  The systems with blueshifts are virtually certain to be cluster members \citep{1978AJ.....83..904S} but the wide range in redshifts could conspire to conceal contaminants.  Indeed, the Virgo W Cluster \citep{1961ApJS....6..213D} is centered slightly to the southwest of the Virgo Cluster in projection, contains galaxies with velocities 1700 $-$ 2700 \kms, but lies at about twice the distance of the Virgo Cluster.  Some galaxies associated with Virgo W can be expected to project onto the Virgo Cluster, including those in the Virgo M structure \citep{1984ApJ...282...19F}.  The Virgo W$^{\prime}$ Group lies at an intermediate distance in the background.  Surface Brightness Fluctuation observations \citep{2009ApJ...694..556B} confirm this picture.  Our own most recent compilation of CosmicFlows-1 distances can be culled to map the structure in the region of the Virgo Cluster \citep{2008ApJ...676..184T}.  Projection contaminants do exist but fortunately all the known problems are confined to the western and southwestern sectors involving $\sim 40\%$ of the area of the cluster.  There are no documented contaminants in the central and eastern sectors \citep{2002MNRAS.335..712T} except for the dwarf GR8 that lies just outside the Local Group.  

Figure~\ref{lbvir} shows the distribution of galaxies at large around the Virgo Cluster and of the specific sample used in the cluster calibration.  The region prone to contamination is identified.  The circle at radius $6.8^{\circ} = 2.0$ Mpc lies at what we infer to be the radius of second turnaround \citep{2010arXiv1010.3787T} which is closely related to the virial radius.  We only accept candidates that project onto the collapsed cluster and in the region free of known contaminants.

The correlation between luminosity and rotation rate was shown in Fig.~\ref{tfivir}.  One $4\sigma$ outlier is identified by the open symbol.  This system, PGC42081=IC3583, is an irregular galaxy with a measured distance assuming the mean relation of only 7 Mpc; ie, well to the foreground.  Plausibly, given its heliocentric velocity of 1121 \kms, this galaxy lies near the foreground zero velocity surface, at the interface between cosmological expansion and cluster infall.  This galaxy is excluded from the Virgo Cluster fit.  From the other 26 candidates a modulus of $31.01\pm0.10$ ($15.9\pm0.8$ Mpc) is determined.  This distance to the Virgo Cluster is consistent with the distance of $16.5\pm1.1$ Mpc favored by \citet{2009ApJ...694..556B} reporting on the Surface Brightness Fluctuation project with Hubble Space Telescope.

\begin{figure}[!]
\includegraphics[scale=0.455]{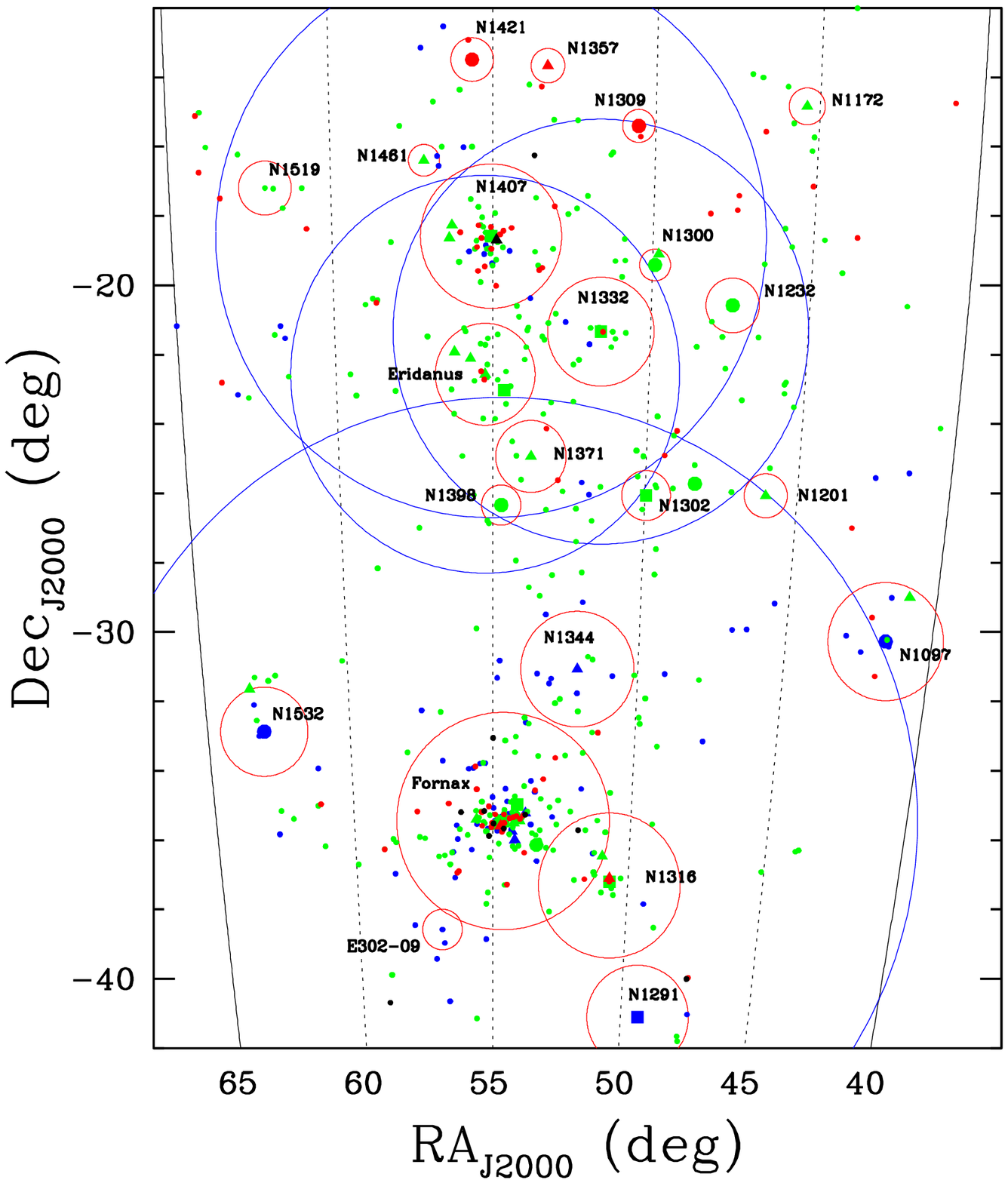}
\includegraphics[scale=0.38]{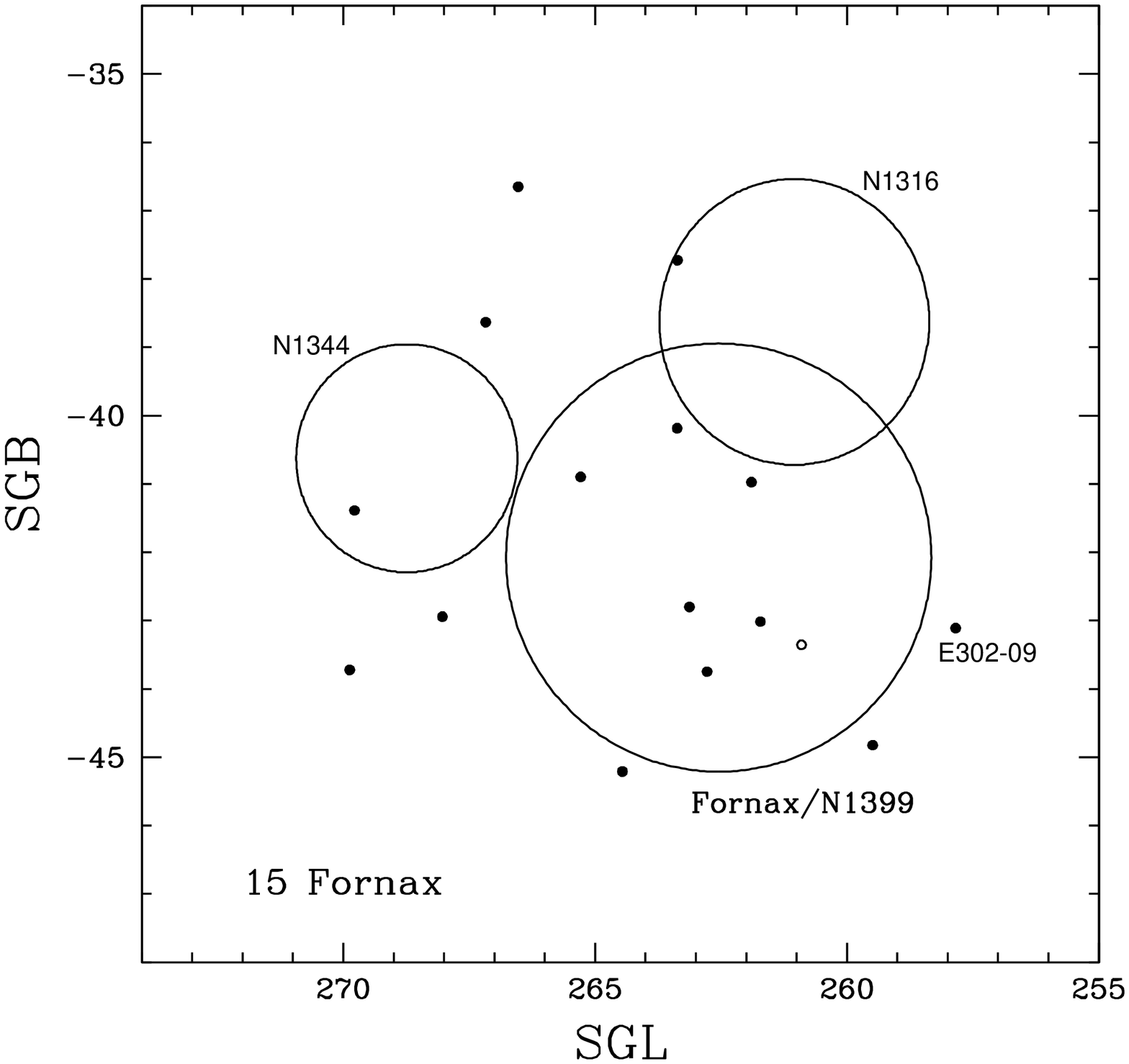}
\caption{{\it Top.} The distribution of galaxies in the Fornax $-$ Eridanus Cloud.  Blue symbols: 800-1300 \kms; green: 1300-1900 \kms; red: 1900-2400 \kms.  Red circles: radii of second turnaround around collapsed halos; blue circles: radii of first turnaround. {\it Bottom.} Galaxies contributing to the cluster calibration sample.  Projection is in supergalactic coordinates and the second turnaround circles from the top panel are superimposed.  The open symbol locates the anomalous galaxy PGC13794=NGC1437B.} 
\label{lbfor}
\end{figure}

\noindent{\bf Fornax Cluster.}  In contrast to the Virgo Cluster, the smaller Fornax Cluster has few spiral galaxies in its core.  It is necessary to accept candidates at somewhat larger distances to accumulate a large enough sample.  In the top panel of Figure~\ref{lbfor} one sees a wide view of the Fornax environment and in the bottom panel there is a zoom into the central region (rotated to supergalactic coordinates) and a display of the useful candidates.

The red and blue circles in the top panel of the figure illustrate (crudely) second and first turnaround surface around the major collapsed regions (halos) in the Fornax$-$Eridanus region.  There is little correlation between velocity and distance across this region and first turnaround surfaces between adjacent halos are strongly overlapping.  There is the implication that the entire region will ultimately collapse.  See \citet{2006MNRAS.369.1351B} for a further discussion.

Several separate halos lie in close proximity to the Fornax Cluster core, taken to center on NGC 1399.  The most important, and overlapping, is centered on NGC 1316.  A third, abutting, is centered on NGC 1344.  The sample of 43 early-type galaxies in the Surface Brightness Fluctuation study by \citet{2009ApJ...694..556B} spills from the NGC 1399 core into the NGC 1316 and NGC 1344 regions.  All three regions are at the same distance to within measurement errors in that study.  Less than half of our candidates lie within the dominant halo identified in Fig.~\ref{lbfor} but, for our purposes, the issue is not whether a candidate lies within the central halo but whether it lies close enough to the others in distance to be statistically representative.  The luminosity-linewidth correlation was shown in Figure~\ref{tf6a}.  The large symbols represent galaxies projected onto the central Fornax halo while the small symbols represent galaxies outside the core.  No statistically significant difference is seen in either component from the fit to the ensemble.  Within the measurement errors all the galaxies in the region are at the same distance, a result consistent with that found by Blakeslee et al.

One galaxy, PGC13794=NGC1437B,  with an excursion of $4\sigma$ is deleted from the fit.  This galaxy has alternatively been classed Sa and I0 and has a strikingly asymmetric HI profile.  The galaxy has probably experienced stripping.

The fit to the remainder of the candidates results in a modulus of $31.19\pm0.11$ ($17.3\pm0.9$ Mpc).  The differential distance modulus with respect to the Virgo Cluster is $0.18\pm0.13$ which is a $2\sigma$ deviation from the \cite{2009ApJ...694..556B} differential modulus of $0.42\pm0.03$.

\noindent{\bf Ursa Major Cluster.}  Perhaps the association of galaxies in Ursa Major should not be called a cluster although it is substantial enough to only fail the Abell richness class zero definition by a couple of galaxies \citep{1996AJ....112.2471T}.  However with galaxies spread across a long axis of over 4 Mpc and a velocity dispersion of only 148~\kms\ the structure involves more than a single parent halo.  Most of the galaxies in the region are gas rich spirals and irregulars with a couple of separated knots of early-type galaxies.  Although contamination could be expected because the entity is in the plane of the Local Supercluster, problems are minimized by the limited velocity window ($700 < V_h +300 {\rm sin} \ell {\rm cos} b < 1210$~\kms) and there is no evidence of interlopers.  For the present discussion, at issue is whether the galaxies in the structure are at similar distances.  We find no differences in distances by location or velocity.  If the structure has a line-of-sight depth comparable to the projected long axis dimension then a dispersion of $\sigma_{los} = 0.17$ mag will contribute to the observed dispersion.

To date, the most ambitious analysis of the luminosity-linewidth correlation in a single cluster was a study with WSRT, the Westerbork Synthesis Radio Telescope \citep{2001A&A...370..765V, 2001ApJ...563..694V}.  The sample in the present study is almost the same as Verheijen's RC (Rotation Curve) sample.  We add six systems that were not observed with WSRT but have single beam profiles (PGC's 36875, 37038, 37832, 38951, 39344, 40537) and remove 3 that are considered confused or morphologically inappropriate (PGC's 35711, 38375, 38643).

\citet{2001ApJ...563..694V} confirmed the finding by \citet{1994AJ....107.1962B} with a Coma Cluster study that with judicious morphological and HI profile preselection (non-interacting, Sb to Sd, steep HI profile edges, smooth outer contours, no bar) the correlation between luminosity and linewidth is much tighter ($\sigma \sim 0.2 - 0.3$) than we find with a larger, less restrictive sample.  When Verheijen gave attention to his RC sample that approximates our own he found two factors in detailed velocity maps that correlate with dispersion.  One of these is found in relatively luminous and earlier typed spirals where rotation curves are observed to peak and then drop to a lower velocity in the outer reaches.  Verheijen found that if a linewidth measure based on a global profile is used (or alternatively, a measure of the maximum velocity) then luminous, early-type galaxies scatter to larger linewidths at a given luminosity but if a linewidth measure indicative of the outer rotation that he calls $V_{flat}$ is used then offsets from the mean relation are removed.  The other correlation with dispersion occurs at the opposite end of the relation, with relatively faint systems, where rotation curves are observed to still be rising at the outermost HI contours.  These galaxies tend to lie to smaller linewidths at a given luminosity, suggesting that the gas is too restricted to sufficiently probe the halo potential.  Both these clarifications concerning the correlation scatter are edifying but cannot be usefully incorporated if only global profiles are available.

Several other conclusions by \citet{2001ApJ...563..694V} are worth noting.  He did not find any convincing second parameters in the near infrared correlation although correlations at $B$ band were found in the sense that early, high surface brightness, red galaxies lie faintward of late, low surface brightness, blue galaxies at a given linewidth.  The color dependence negates the offsets in the infrared.  Verheijen also confirmed the improvement to the correlation found by   \citet{2000ApJ...533L..99M} if luminosity is replaced with a measure that sums the mass in stars and the mass in gas (the baryonic mass).

In summary of this discussion, the spatially resolved kinematic study by \citet{2001ApJ...563..694V} clarifies several issues but these go beyond what is useful for our present purposes.  We only have access to global profiles and we want a methodology that minimizes subjective restrictions based on morphology.  The methodology should be as inclusive as possible and not have characteristics that might be interpreted differently as a function of distance.

The luminosity-linewidth correlation for the Ursa Major association is shown in Figure~\ref{tf6a} and the distance modulus determined from 34 galaxies in that correlation is $31.20\pm0.10$ ($17.4\pm0.9$ Mpc).  By contrast, the distance modulus to 7 early-type galaxies in Ursa Major from Surface Brightness Fluctuation measurements \citep{2001ApJ...546..681T} is $30.69\pm0.10$ ($13.7\pm0.7$ Mpc), a difference with $3.6\sigma$ significance from our value.  Overall around the sky there is agreement between surface brightness fluctuation and luminosity-linewidth measures \citep{2008ApJ...676..184T} but this significant difference in Ursa Major remains an abiding mystery.

\begin{figure}[!]
\includegraphics[scale=0.48]{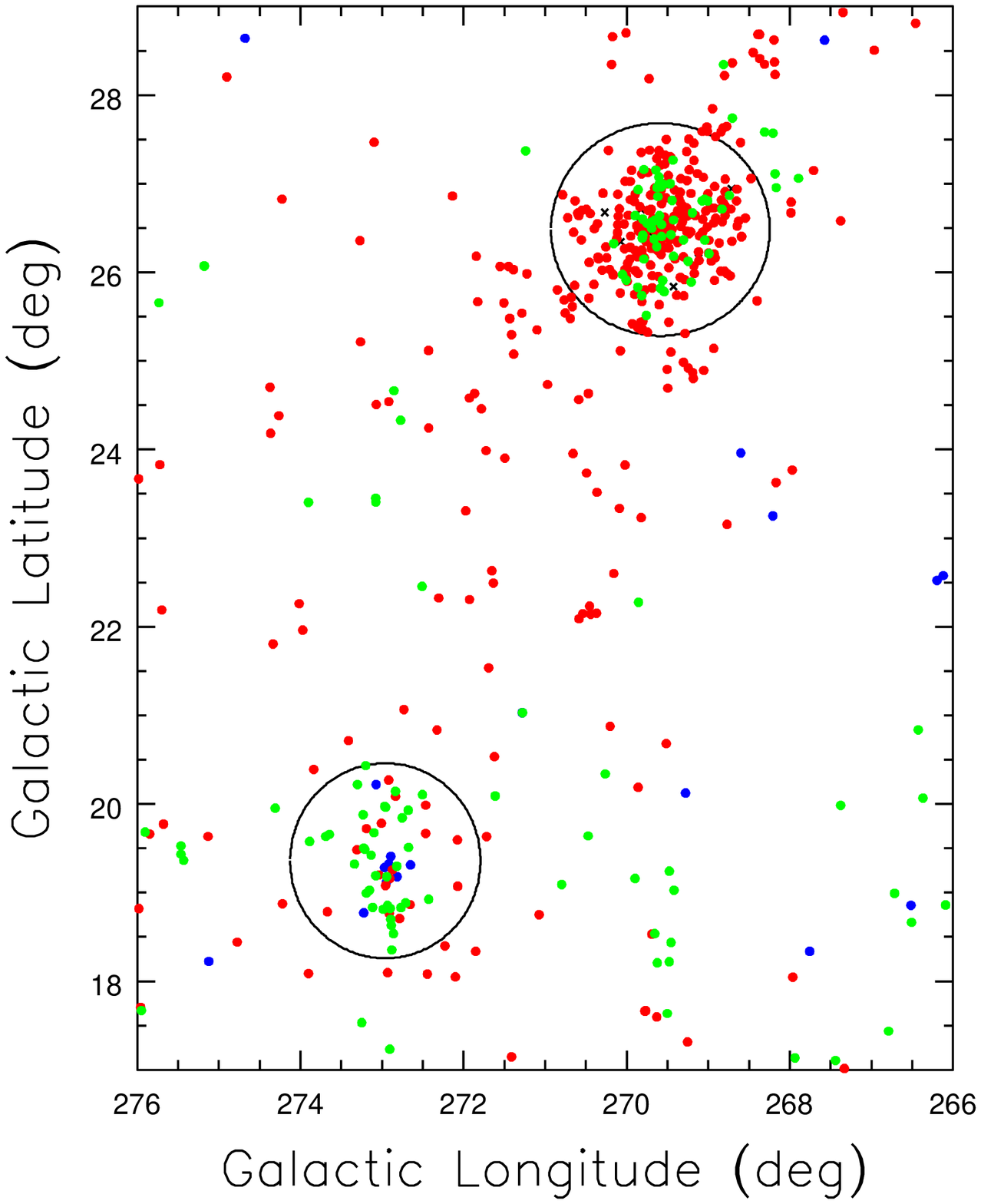}
\includegraphics[scale=0.41]{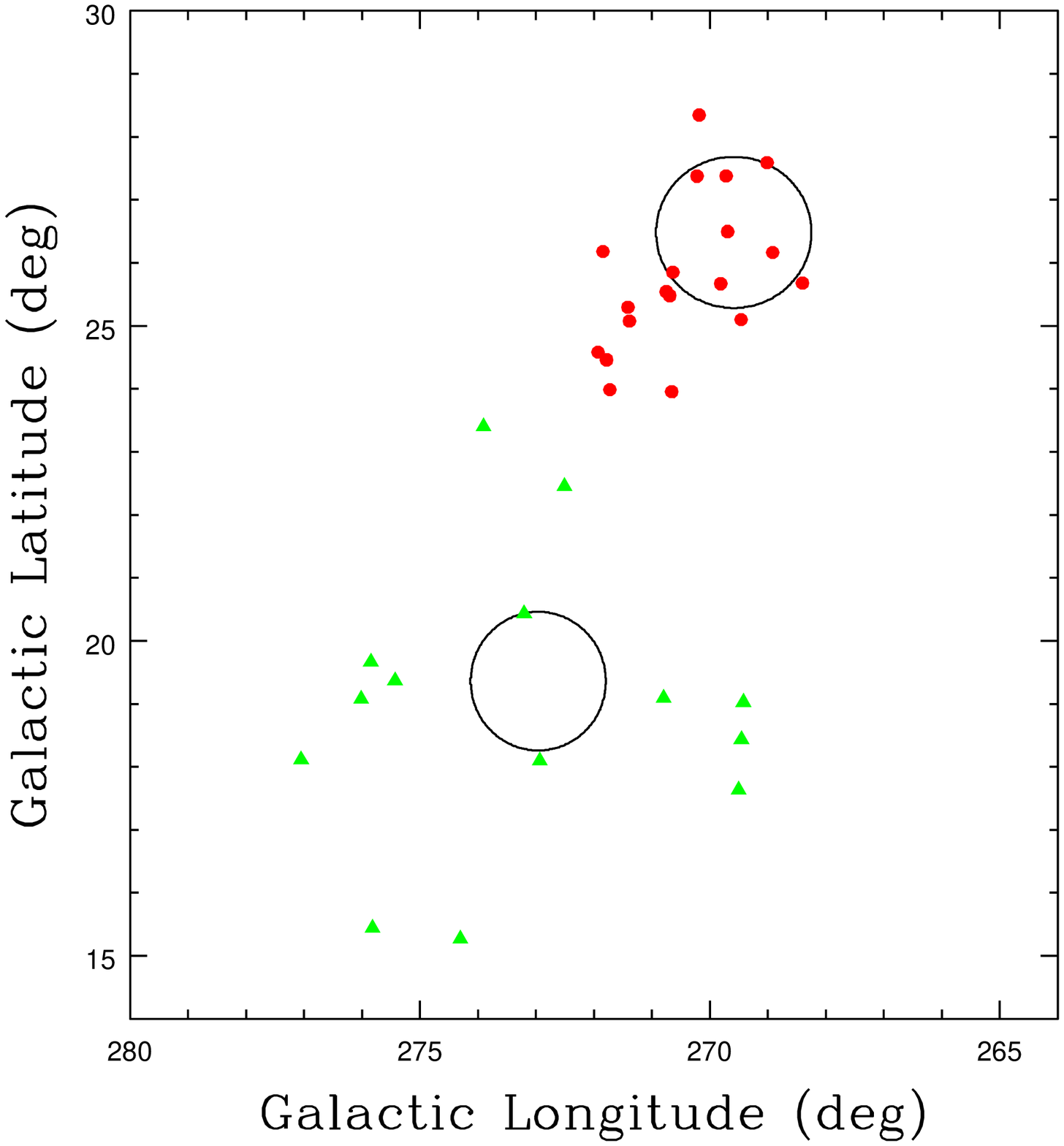}
\caption{{\it Top.}  Galaxies in the region of the Antlia and Hydra clusters.  Blue: $V_h < 2000$ \kms; green: 2000-3000 \kms; red: 3000-5000 \kms. Antlia Cluster lies within the lower left circle and Hydra Cluster lies within the upper right circle. {\it Bottom.} The Antlia calibration sample distribution is shown by the green triangles while the Hydra calibration sample is located by the red circles.} 
\label{lbhyan}
\end{figure}

\noindent{\bf Antlia and Hydra Clusters.}  These two clusters are treated together because, although they are at quite different distances, they lie close to each other on the sky and their velocity ranges overlap.  Confusion is possible.  The top panel of Figure~\ref{lbhyan} shows the distribution of galaxies in the neighborhood with systemic velocities indicated by colors and the collapsed cores of the two clusters shown by circles: Antlia to the lower left with smaller velocities and Hydra to the upper right with larger velocities.  In the bottom panel of this figure the green triangles and red circles respectively locate the galaxies that constitute the Antlia and Hydra samples.  In both cases, the samples spill outside the core regions, extremely so in the case of the Antlia Cluster.  Both clusters are dominated by early-type systems in the collapsed cores.  Two of the Antlia candidates lie in the direction of the Hydra Cluster and might be a particular concern but both are more consistent in velocity and measured distance with membership in Antlia.  The luminosity-linewidth diagram for these two clusters are seen in Figure~\ref{tf6a}.  The scatter in the Antlia plot is particularly small, minimizing concerns about depth variations though the sample is scattered on the sky.  By contrast the spatially more compact Hydra sample manifests the greatest correlation scatter of any of the clusters.  Still, the slope derived from Hydra Cluster alone is in excellent agreement with that of the 13 cluster ensemble.

The Antlia Cluster is determined to have a distance modulus of $32.85\pm0.10$ ($37.2\pm1.7$ Mpc).  This value is in good agreement with the average of 3 galaxies in Antlia with Surface Brightness Fluctuation measurements \citep{2001ApJ...546..681T} of $32.64\pm0.14$ ($33.7\pm2.4$ Mpc).  The modulus of the Hydra Cluster is $33.84\pm0.14$ ($59\pm4$ Mpc).  This result is in poor agreement with the Surface Brightness Fluctuation measure by \citet{2005A&A...438..103M} of $33.07\pm0.07$ ($41.2\pm1.4$ Mpc), a $5\sigma$ excursion from our value.

\begin{figure}[!]
\includegraphics[scale=0.39]{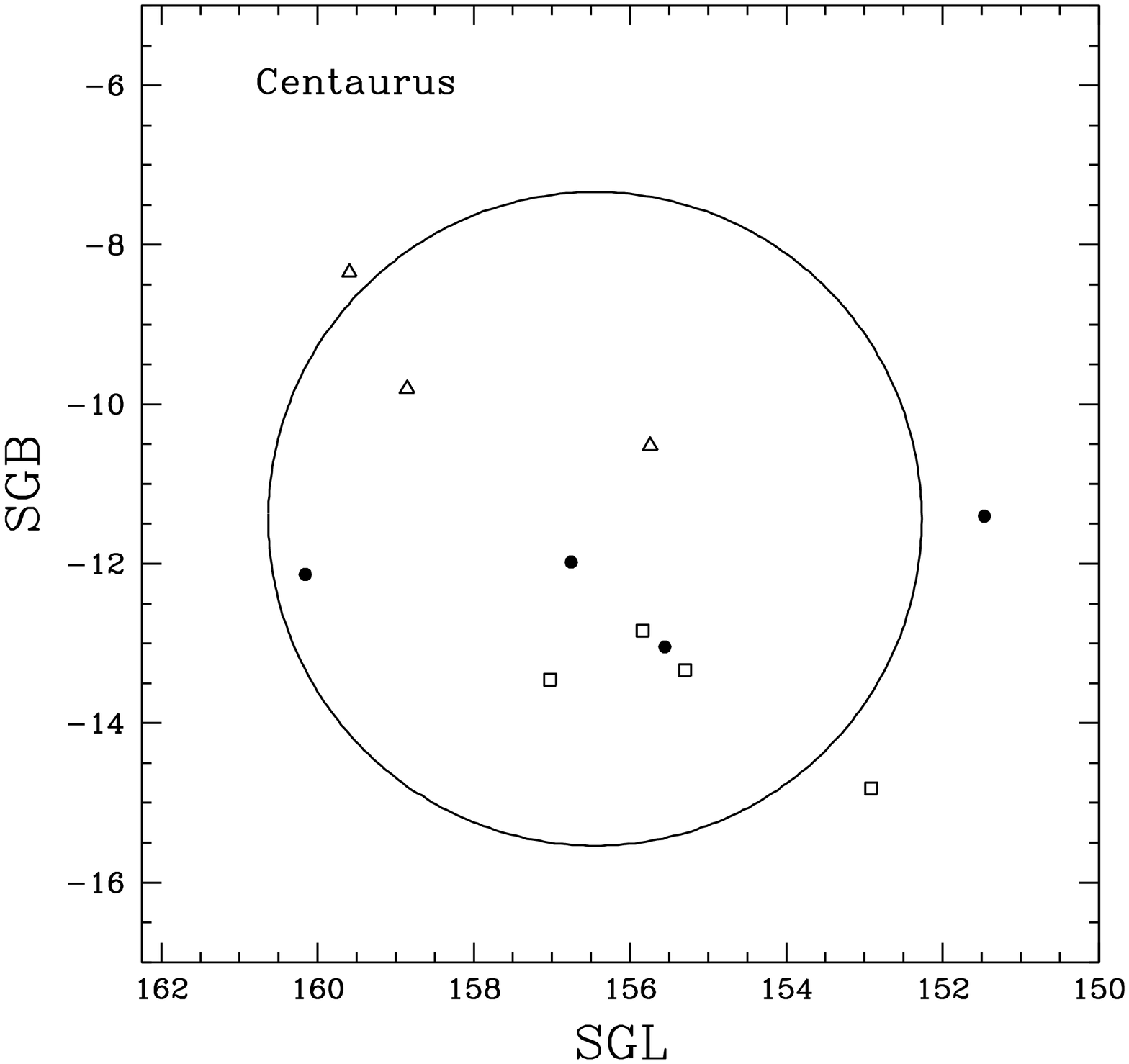}
\includegraphics[scale=0.38]{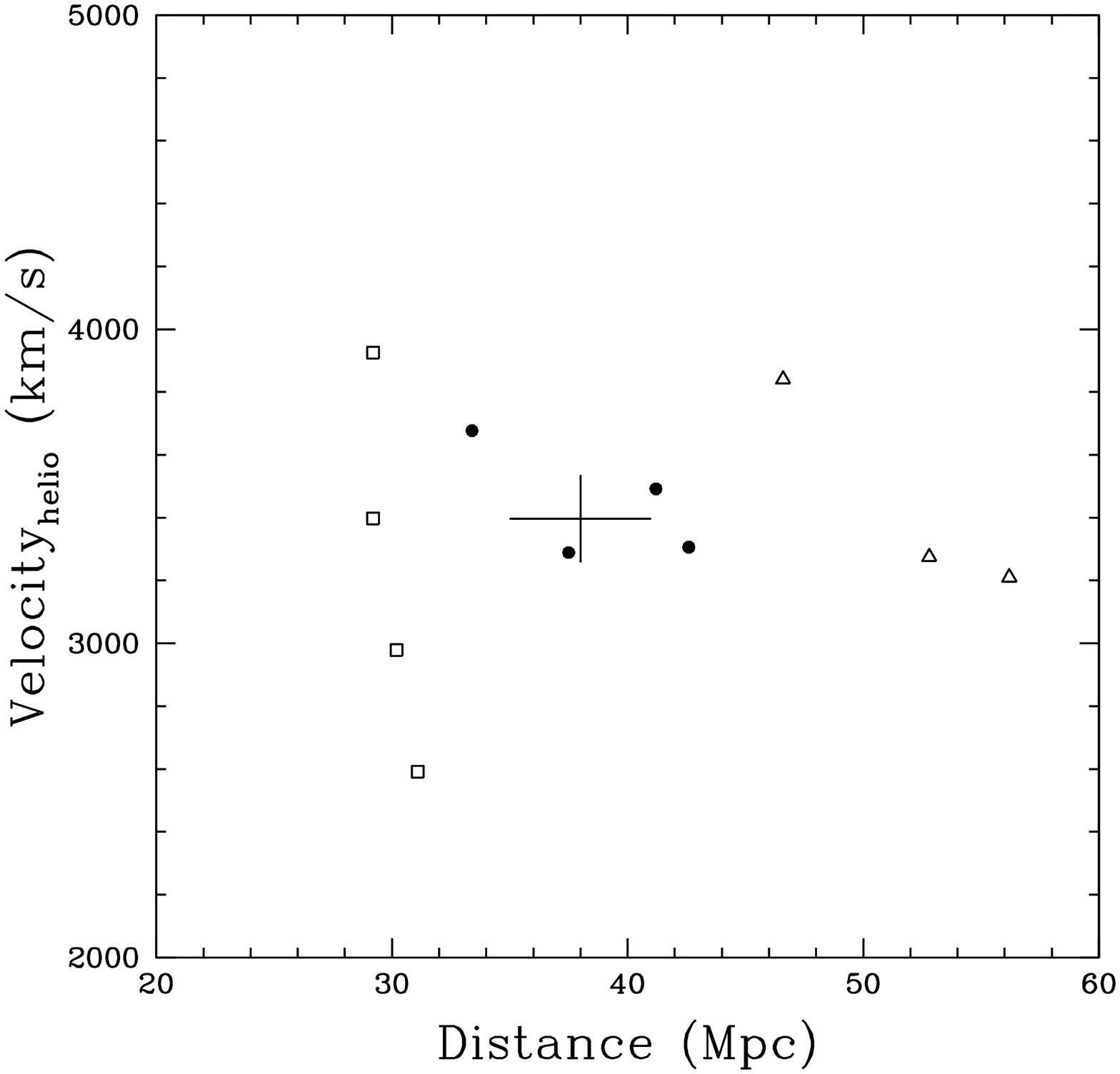}
\caption{{\it Top.} Distribution of Centaurus sample. The second turnaround radius of the collapsed core lies within the circle. Galaxies  that lie greater than 0.4 mag above and below the mean luminosity-linewidth correlation are identified by open squares and triangles respectively. {\it Bottom.} Dependence of velocities on inferred distances assuming candidates intrinsically obey the mean luminosity-linewidth relationship.  Symbols as in panel above.  The mean distance and velocity of the cluster is indicated by the cross.  The arms of the cross indicate $1\sigma$ limits.} 
\label{lbcen}
\end{figure}

\noindent{\bf Centaurus Cluster.}  Two distinct kinematic structures in the line of sight complicate the situation with this cluster.  The complexity is described by \citet{1997A&A...327..952S}.  In a histogram of velocities there is a narrow peak at 4750~\kms\ superposed on a broad distribution centered at 3400~\kms.  The spatial centroid of the higher velocity component is displaced on the sky from the centroid of the broad distribution.  The two components have come to be referred to as the high velocity Cen45 and the low velocity Cen30 \citep{1986MNRAS.221..453L}. Alternatively, the two features could be at distinct distances in projection \citep{1989ApJS...69..763F} or in the process of merging, hence at similar distances \citep{1988MNRAS.235.1177L}.

The current effort is to calibrate a way to measure distances so we want to avoid controversy that will be resolved by measuring distances.  Consequently we impose a restriction on velocities for this sample, $V_h < 4000$~\kms, that excludes any Cen45 contribution.

The luminosity-linewidth correlation found with the resultant sample is displayed in Fig.~\ref{tf6a}.  The scatter is uncharacteristically large.  Figure~\ref{lbcen} shows the sky projection of candidates.  They lie within or very close to the cluster core defined by an estimate of the second turnaround collapse radius.  There is the curiosity that the most deviant cases brighter than the mean relation ($\Delta \mu > 0.4$) lie on one side of the cluster (at more positive supergalactic latitudes) while the most deviant in the opposite sense ($\Delta \mu < -0.4$) lie on the other side.  There might be a hint here of a depth effect, but the lower panel of Fig.~\ref{lbcen} provides no support.  In this plot, scatter from the mean relation is interpreted as distance offsets which are plotted against systematic velocity but there is no suggestion of the S$-$wave of infall that would be anticipated.  Nor is there a hint of the Hubble relation with distance.  Our conclusion is to take the sample at face value, as representative of the Cen30 cluster with unexplained enhanced scatter.  The data supports a distance modulus of $32.91\pm0.16$ ($38\pm3$ Mpc).  By comparison \citet{2005A&A...438..103M} found $33.28\pm0.09$ ($45\pm2$ Mpc) from a surface brightness fluctuation study.  The difference is $0.37\pm0.18$, roughly $2\sigma$.  Most perplexing, though, is that we find Centaurus Cluster to lie $0.93\pm0.21$ convincingly {\it in front of} Hydra Cluster while Mieske et al. claim Centaurus is $0.21\pm0.11$ {\it behind} Hydra.

\begin{figure}[t]
\includegraphics[scale=0.39]{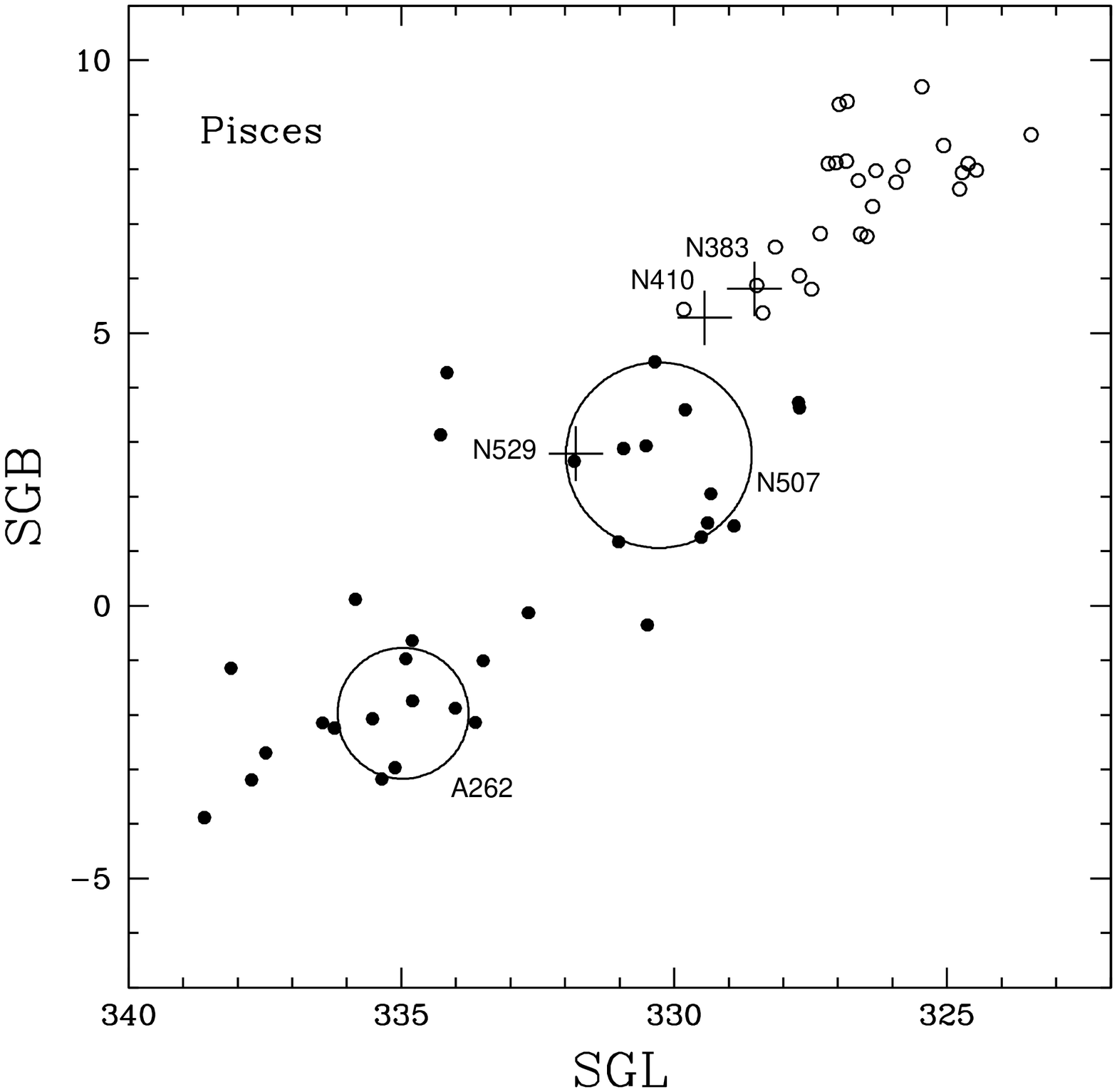}
\caption{Galaxies in the Pisces filament sample in supergalactic coordinates.  The location of the NGC 507 Cluster is given by the large circle, that of Abell 262 is given by the small circle, and other suggested groupings are indicated by crosses.  Members of the sample at SGB $<5$ are shown as filled symbols and those at SGB $>5$ are shown as open symbols.}
\label{lbpis}
\end{figure}

\begin{deluxetable}{lccc}
\tablenum{A1}
\tablecaption{Pisces Filament Sub-samples}
\label{tbl:pis_sub}
\tablewidth{0in}
\tablehead{\colhead{Sample} & \colhead{Modulus} & \colhead{R.M.S.} & \colhead{No.}}
\startdata
SGB $>5$ & $34.07\pm0.10$ & $\pm0.53$ & 26 \\
NGC 507  & $33.92\pm0.08$ & $\pm0.31$ & 15 \\
A 262         & $34.03\pm0.09$ & $\pm0.37$ & 17 \\
SGB $<5$ & $33.98\pm0.06$ & $\pm0.34$ & 32 \\
\hline
All               & $34.02\pm0.06$ & $\pm0.43$ & 58 \\
\enddata
\end{deluxetable}

\noindent{\bf Pisces Filament.}  If there is doubt that the Ursa Major region should be considered a cluster, it is certain that the region we will consider in the well studied Perseus$-$Pisces Supercluster \citep{1993AJ....105.1251W} is not just a cluster.  It can be seen from the distribution of the sample in Figure~\ref{lbpis} that galaxies are being considered over a projected long axis extent of 20 Mpc.  There are several knots dominated by early-type systems (so not represented in our figure), most prominently a cluster around NGC 507 and the Abell 262 Cluster.  The inferred second turnaround cores of these two entities are indicated in Fig.~\ref{lbpis} along with the locations of several smaller groupings \citep{1994AJ....108...33S}.

The region shares two characteristics with Ursa Major that make it interesting for us.  First, in both cases the structures are evidently aligned almost in the plane of the sky so dispersion in depth does not contribute much to the dispersion of the luminosity-linewidth relation.  Of course, this assertion has to be checked.  Second, the galaxies in the structures are predominantly gas-rich spirals in uncrowded environments.  Most galaxies that we will ultimately want to consider are in similar environments.

It had already come to our attention with the calibration of Cosmicflows-1 that the dispersion along a segment of the Pisces filament is small and implied unmeasurable depth \citep{2000ApJ...533..744T}.   \citet{1997MNRAS.291..488H} acquired distances along a larger segment of the Perseus$-$Pisces chain using the Fundamental Plane methodology and, while they found variations in distance along the full length, they could not see any distance variations along the length that interests us.

In an initial fit, all 58 galaxies in the sample are considered.  The slope fit to this sample alone is in good agreement with the 13 cluster `universal' slope.  A fit constrained to the universal slope and with a zero point established by the local calibrators results in a distance modulus of $34.02\pm0.06$ ($64\pm2$ Mpc).  The rms scatter is $\pm0.43$.

It was noticed, though, that the most deviant points lie at the north end of the filament (in supergalactic coordinates).  To test for variations along the filament we split the sample into three parts: (i) SGB $>5$, (ii) nearest NGC 507, and (iii) nearest A 262.  The mean modulus, dispersion, and sample size for the three parts and for the ensemble are summarized in Table~\ref{tbl:pis_sub}. The results are clear.  There are no significant differences between any of the moduli.  However the scatter is much greater in the subsample at SGB $>5$; rms scatter of $\pm0.53$ in magnitude for 26 galaxies compared with $\pm0.34$ for 32 galaxies at SGB $<5$.  The scatter in the southern subsample is amazingly low, leaving little place for depth effects.  By contrast the scatter in the northern subsample is large and an obvious explanation would be a contribution from distance variations.  Nevertheless, the fit to the ensemble of 58 galaxies discussed in the previous paragraph is representative within certainties of all the subsamples and the rms scatter for the ensemble, averaged over the exceptional south and the spotty north, is about normal.  The luminosity-linewidth relation is seen in Fig.~\ref{tf6b}.

\noindent{\bf Coma Cluster.}  This cluster merits special attention.  While the Ursa Major and Pisces regions are representative of moderate density environments dominated by gas-rich galaxies, Coma Cluster represents the high density extreme, a place dominated by gas-poor systems.  Observations of the sample present compounded difficulties: the targets are faint in HI flux both because they are far away and intrinsically.  For the clusters that have been discussed up to now HI detections are generally not a challenge.  With the Coma sample longer integrations were sometimes required to obtain adequate spectra.  With considerable effort, essentially all our original candidates are usefully detected.

As an extra motivation, \citet{1994AJ....107.1962B} have reported that the dispersion in the luminosity-linewidth relation is very small in Coma.  Is the slope of the correlation consistent with that found in other clusters?

Our version of the relation is shown in Figure~\ref{tf6b}.  Indeed, the dispersion is small (rms of $\pm0.27$ mag) about a slope to this sample of $-6.96\pm0.56$.  The difference between the two slopes in the figure has a statistical significance of $3\sigma$.  It is seen in Figure~\ref{slopes} that all the other individual cluster slopes are within $\sim 1\sigma$ of the mean slope.  In the case of Coma the error bar is small because the scatter around the shallow slope is small.

\begin{figure}
\includegraphics[scale=0.9]{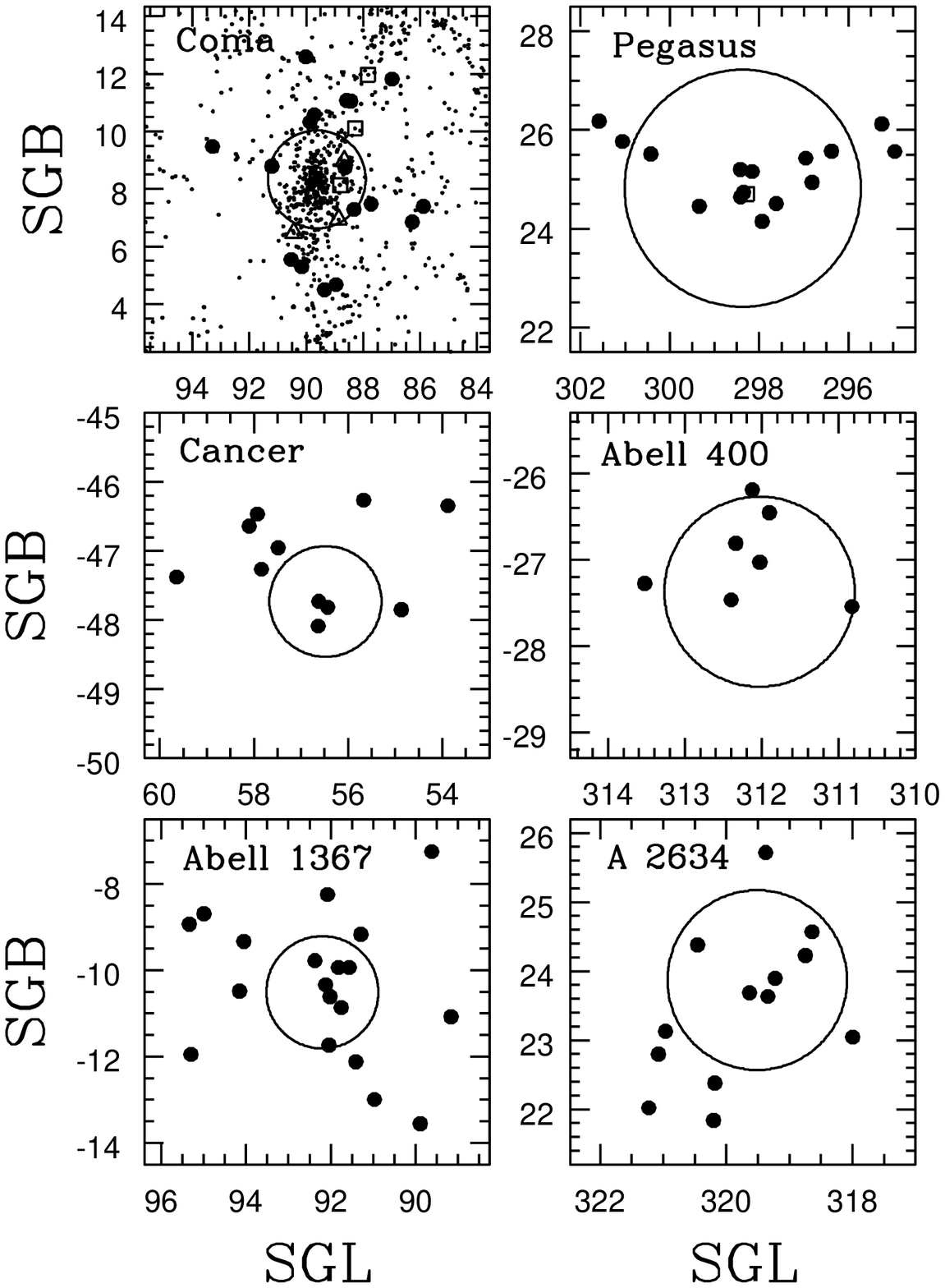}
\caption{Projection of six samples in supergalactic coordinates.  In all the panels the large circles approximate the second turnaround radii of the collapsed cores.  In the case of the Coma sample the small dots show the distribution of a redshift sample while the large symbols identify the galaxies used in the template:  open squares: cases that scatter $>0.4$ mag brightward of the mean relation; open triangles: cases that scatter $>0.4$ mag faintward.  In the case of the Pegasus Cluster the open square identifies the location of PGC 71159, the most strongly deviant galaxy from the linewidth correlation.}
\label{lb6}
\end{figure}

Figure~\ref{lb6} shows the locations of the sample in relation to the cluster.  The sample candidates are not very concentrated.  Only 6 of 23 lie within the region of collapse.  The open symbols identify systems that deviate by more than 0.4 mag from the fit with the universal slope; squares to the bright side and triangles to the faint side.  It is seen that the deviant cases preferentially lie closest to the cluster core.  Five of the six open symbols are within or adjacent to the boundary of the core.  The two largest deviants ($\sim 0.8$ mag) are within the circle defining the core and one of these is the candidate closest to the center (PGC 44416).

At issue is whether the Coma relation is actually different $-$ shallower slope and less dispersion $-$ or whether the relation is the same as elsewhere and the differences result from statistical happenstance.  We see from Fig.~\ref{slopes} that the slope difference is (at the outer limits) within the range of differences found with other cluster samples and recognize that the $3\sigma$ measure of departure is a consequence of the anomalously small scatter.  Accepting the universal slope, the rms scatter is $\pm0.38$ which is normal.  We note that the most deviant cases driving the flatter slope are within the cluster core.  The most straight forward conclusion is that the Coma sample is drawn from the universal relation.  On that basis, the distance modulus is determined to be $34.77\pm0.08$ ($90\pm4$ Mpc).  

\noindent{\bf Pegasus, Cancer, A400, A1367, and A2634 Clusters.}  The five remaining clusters are all legacies of early work by the Aaronson, Huchra, Mould collaboration \citep{1985ApJS...57..423B}.  The clusters all lie in the declination range ($-1$ to $+38$) accessible to the Arecibo radio telescope, the most sensitive single-aperture facility for HI line studies.

In the case of the Pegasus Cluster, the one $2\sigma$ outlier (PGC 71159) is seen in Figure~\ref{lb6} to lie at the very center of the cluster, a circumstance that might suggest an anomaly.  In the case of Cancer, \citet{1983ApJ...268...47B} have convincingly shown that this region, like with Pisces and Ursa Major, is more complex than just a cluster.  The circle in Figure~\ref{lb6} shows the location of the main concentration of galaxies identified from a redshift survey.  Our sample is drawn from a more extended area.  Regarding A400, A1367, and A2634 there is not much to be said.   Figure~\ref{lb6}, shows the sky distributions in each case.  The scatter about the central cluster is somewhat greater with A1367 and reasonably confined with the other two.  There are no anomalies in the luminosity-linewidth plots seen in Figure~\ref{tf6b}.  A2634 is noteworthy as the significantly most distant cluster in this collection.

\onecolumn
\twocolumn





\bibliography{paper}
\bibliographystyle{apj}

\
\end{document}